\documentclass[journal,10pt]{IEEEtran}
\pdfoutput=1
\IEEEoverridecommandlockouts

\usepackage[utf8]{inputenc} 
\usepackage[T1]{fontenc}
\usepackage{url}
\usepackage{ifthen}
\usepackage{cite}
\usepackage[cmex10]{amsmath} % Use the [cmex10] option to ensure complicance
\usepackage{amsmath,amssymb,amsfonts}
\usepackage{algorithmic}
\usepackage{graphicx}
\usepackage{textcomp}
\usepackage{xcolor}
\usepackage{amssymb}
\usepackage{graphicx}
\usepackage{epstopdf}
\usepackage{balance}
\usepackage{epsfig}
\usepackage{graphics}
\usepackage{subfigure}
\usepackage{lineno}
\usepackage{textcomp}
\usepackage{bm}
\usepackage{upgreek}
\usepackage{filecontents}
\usepackage{setspace}
\usepackage{hyperref}
\usepackage[ruled,linesnumbered]{algorithm2e}
\usepackage{acronym}

\usepackage{acronym}

\acrodef{cas}[CAS]{communication-assisted sensing}
\acrodef{snc}[S\&C]{sensing and communication}
\acrodef{rd}[R-D]{rate-distortion}
\acrodef{iid}[i.i.d.]{independently and identically distributed}
\acrodef{mse}[MSE]{mean squared error}
\acrodef{mmse}[MMSE]{minimum mean squared error}
\acrodef{glm}[GLM]{Gaussian linear model}
\acrodef{wrt}[w.r.t.]{with respect to}
\acrodef{sct}[SCT]{source-channel separation theorem}
\acrodef{qos}[QoS]{quality of service}
\acrodef{snr}[SNR]{signal to noise ratio}
\acrodef{sw}[SW]{separated S\&C waveforms}
\acrodef{dw}[DW]{dual-functional waveform}
\acrodef{trm}[TRM]{target response matrix}
\acrodef{sdas}[SDAS]{simultaneous sensing data acquisition and sharing}

\newcommand{\Rv}[1]{\mathsf{#1}}

\newcommand{\Prob}[2]{P_{\mathsf{#1}}(#2)}

\newcommand{\Claw}[2]{Q_{\mathsf{#1}}(#2)}

\newcommand{\normquad}[1]{\left\lVert #1\right\rVert_2^2}

\newcommand{\opmin}[1]{\mathop{\min}\limits_{#1}}

\newcommand{\opmax}[1]{\mathop{\max}\limits_{#1}}

\newcommand{\mathept}[1]{\mathbb{E}\left[ #1 \right]}

\newcommand{\tr}[1]{\text{Tr}\left( #1 \right)}

\graphicspath{{Figures/}}
\begin{document}
	
\title{Simultaneous Sensing Data Acquisition and Sharing in Low-Altitude Wireless Networks: Fundamental Limits and Optimal Signaling} 

\author{Fuwang Dong,~\IEEEmembership{Member,~IEEE}, Fan Liu,~\IEEEmembership{Senior Member,~IEEE}, Yifeng Xiong,~\IEEEmembership{Member,~IEEE}, \\ Yuanhao Cui,~\IEEEmembership{Member,~IEEE}, Wei Wang,~\IEEEmembership{Senior Member,~IEEE}, Shi Jin,~\IEEEmembership{Fellow,~IEEE} 

\thanks{(\textit{Corresponding author: Fan Liu.})}
\thanks{Part of this paper was presented at IEEE International Symposium on Information Theory (ISIT), 2024 \cite{dong2024fundamental}.}
%\thanks{This work was supported in part by the National Key R\&D Program of China (No. 2021YFB2900200), in part by the National Natural Science Foundation of China under Grant 62101234 and Grant U20B2039, and in part by the Young Elite Scientist Sponsorship Program by the China Association for Science and Technology (CAST) under Grant No. YESS20210055. }
\thanks{Fuwang Dong, and Wei Wang are with the College of Intelligent System Science and Engineering, Harbin Engineering University, Harbin 150001, China. (email: \{dongfuwang, wangwei407\}@hrbeu.edu.cn).}

\thanks{Fan Liu, and Shi Jin are with the National Mobile Communications Research Laboratory, Southeast University, Nanjing 210096, China. (e-mail: \{fan.liu, jinshi\}@seu.edu.cn).}

\thanks{Yifeng Xiong, and Yuanhao Cui are with the School of Information and Electronic Engineering, Beijing University of Posts and Telecommunications, Beijing 100876, China. (email: \{yifengxiong, yuanhao.cui\}@bupt.edu.cn).} 

}

\maketitle

\begin{abstract}
In the low-altitude wireless networks, the simultaneous sensing data acquisition and sharing (SDAS) through an ISAC signaling strategy becomes a typical application scenario. In this paper, we mainly investigate three primary aspects of the SDAS system, namely, the information-theoretic framework, the optimal distribution of channel input, and the optimal waveform design for Gaussian signaling. First, we establish the information-theoretic framework and develop a modified source-channel separation theorem (MSST) tailored for the SDAS systems. The proposed MSST elucidates the relationship between achievable distortion, coding rate, and communication channel capacity in cases where the distortion metric is separable for sensing and communication (S\&C) processes. Second, we present an optimal channel input design for dual-functional signaling, which aims to minimize SDAS distortion under the constraints of the MSST and resource budget. We then conceive a two-step Blahut-Arimoto (BA)-based optimal search algorithm to numerically solve the functional optimization problem. Third, to provide practical design insights, we further propose an optimal waveform design for Gaussian signaling in multi-input multi-output (MIMO) SDAS systems. The associated covariance matrix optimization problem is addressed using a successive convex approximation (SCA)-based waveform design algorithm. Finally, we provide numerical simulation results to demonstrate the effectiveness of the proposed algorithms, which characterizes the unique performance tradeoff between S\&C processes.             
                       
\end{abstract}


\section{Introduction}\label{intro}

Unmanned aerial vehicles (UAVs) play a pivotal role in low-altitude wireless networks (LAWN), serving as temporary communication base stations (BS) or aerial monitoring platforms by utilizing their high mobility \cite{wu2025low}. Recently, the integrated sensing and communication (ISAC) technique has revolutionized UAV systems by offering notable advantages such as mutually beneficial interactions between \ac{snc}. The favorable attributes significantly broaden the potential applications of ISAC-empowered UAV, including smart transportation and agriculture \cite{10004900,10098686}. In many scenarios, UAVs are required for target sensing and subsequently transmitting the observed parameters to remote users, especially when direct line-of-sight (LoS) sensing channels between the users and targets are blocked. Thus, the performance is typically quantified by the error between the ground truth and the user's recovery. This closely resembles the established \textit{Remote Estimation} problem \cite{6286997,8635889,gao2018optimal,7935515} when the \ac{snc} processes are independent.

However, a novel paradigm, which we term \ac{sdas} in this paper, emerges when UAVs leverage dual-functional ISAC signals to simultaneously sense current target states and convey previously acquired state information to end-users. The \ac{snc} processes are deeply coupled through the transmitted signal, leading to fundamental differences in both theoretical model and signal design. In what follows, we discuss the key differences between our work and related works from both an application layer and an information-theoretic perspective.

\begin{figure}[!t]
	\centering
	\includegraphics[width=3.5in]{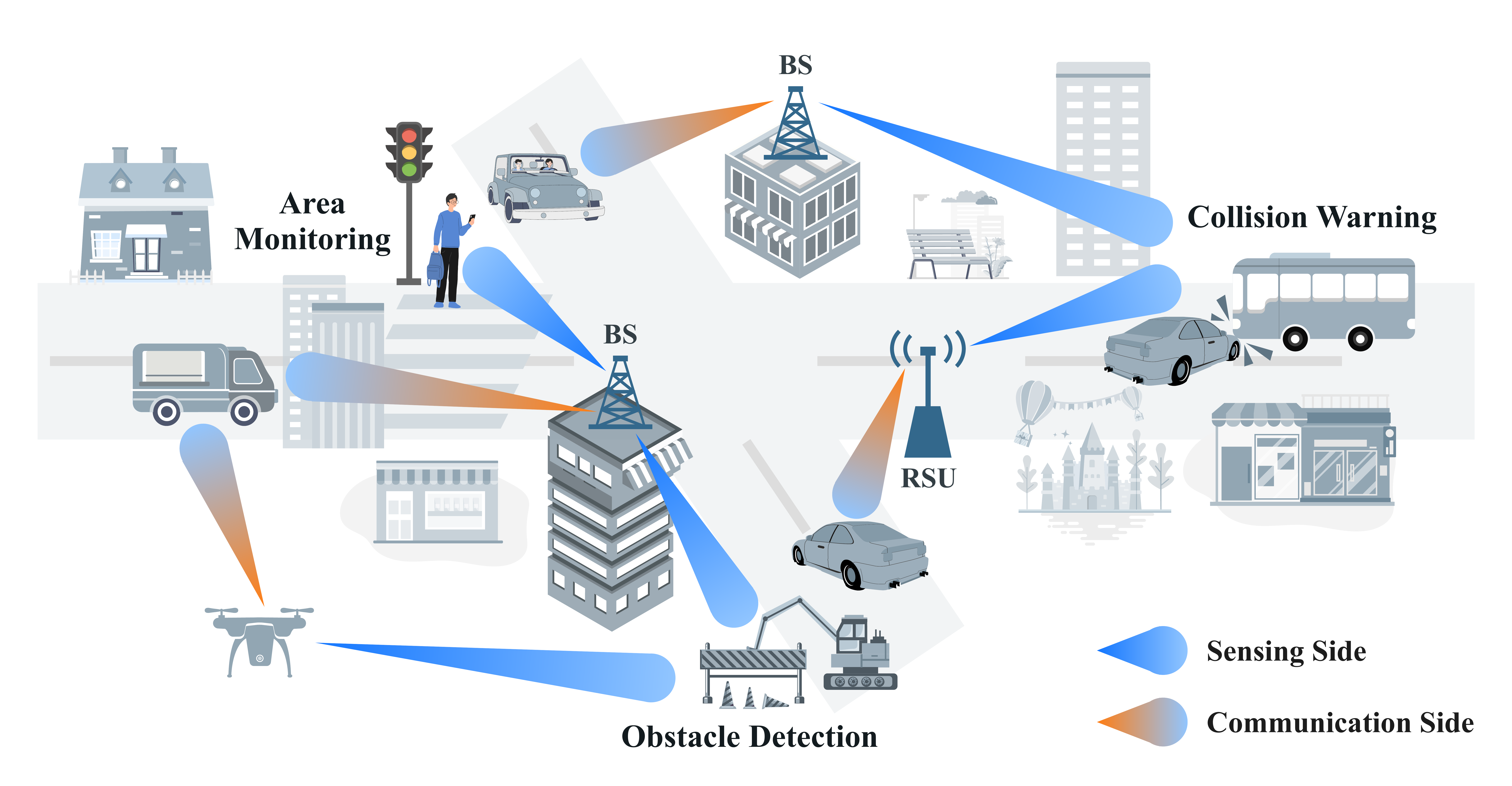}
	\caption{The use case of \ac{sdas} systems in LAMN.}
	\label{CAS_illustration}
\end{figure}

\subsection{Related Works on Application Layer}

\subsubsection{Traditional ISAC Systems}
Most research on ISAC-UAV primarily focus on the joint optimization of trajectory planning, resource allocation and waveform design. These studies often consider three key performance indicators: sensing metrics (e.g., signal to interference and noise ratio (SINR), Cramér-Rao Bound) \cite{10529184}, communication metrics (e.g., achievable rate) \cite{9916163,9858656} and energy consumption \cite{10680299}. Typically, one of these (or a weighted sum) is chosen as the optimization objective, with the others imposed as constraints. Evidently, \ac{snc} performance are characterized through individual and even incompatible evaluation system, e.g., sensing for estimation theory and communication for information theory. However, the \ac{sdas} framework is quantified by the error between ground truth and user's recovery, implicitly reflecting the integrated impact of the \ac{snc} processes. To our best knowledge, the \textit{\ac{sdas} distortion} has not been well-defined, and its minimization design remains largely unexplored.      

\subsubsection{Relay Systems}
Relay systems primarily function by forwarding the signals from a source transmitter to a destination receiver, typically operating in the following two modes: decode-and-forward the signal or simply amplify-and-forward the signal \cite{4273702,4251177}. Despite its structure is similar to that of \ac{sdas}, its main purpose is to transmit communication signals in a cooperative manner without actively performing sensing functions. Furthermore, channel estimation in the relay systems employs either a cooperative training method \cite{5288504} or dividing the estimation into individual source-relay and relay-destination channels \cite{6203610}. However, a quantitative analysis of how the \ac{snc} coupling process affects the \ac{sdas} distortion is still lacking. 

\subsubsection{Communication-assisted Sensing Systems}
In previous work \cite{10233679,10845869}, the authors established a communication-assisted sensing framework and proposed \ac{sdas} distortion minimization-based waveform designs for separated \ac{snc} and dual-functional waveform schemes, respectively. However, the information-theoretic framework of \ac{sdas} has not been rigorously validated from a coding theory perspective, remaining an unclear operational meaning. Moreover, the reliance on Gaussian distribution for dual-functional ISAC waveform design in \cite{10845869} poses a potential sub-optimal property for the \ac{sdas} scenario, especially considering the fact that Gaussian distribution is communication-optimal under Gaussian channel, whereas constant-modulus modulation is sensing-optimal \cite{8437621}.

The above research gap motivates us to delve into the fundamental limits and the optimal waveform design of the dual-functional ISAC signaling strategy for the \ac{sdas} systems.

\begin{figure}[!t]
	\centering
	\includegraphics[width=3in]{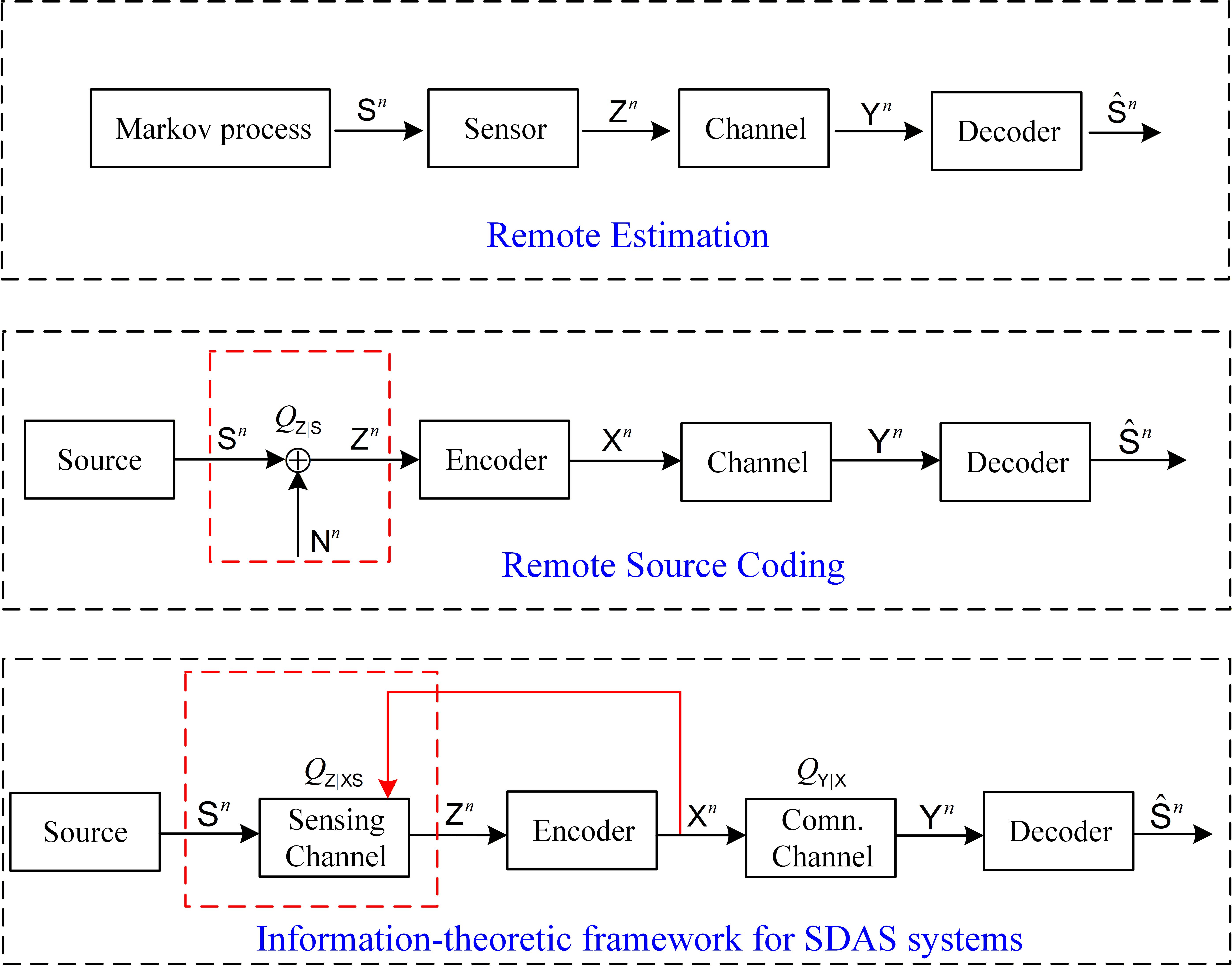}
	\caption{The comparisons of the information-theoretic frameworks between the remote estimation, remote source coding, and the \ac{sdas} system.}
	\label{fig2}
\end{figure}

\subsection{Related Works on Information-theoretic Framework} \label{relatework}

The pioneering studies \cite{8437621,9785593} have considered a scenario where the transmitter (Tx) communicates with a user through a memoryless state-dependent channel while simultaneously estimating the state from generalized feedback. The capacity-distortion-cost tradeoff of this channel is characterized to illustrate the optimal achievable rate for reliable communication while maintaining a preset state estimation distortion. The authors of \cite{11155173} further extend the capacity-distortion function to a framework based on backward simultaneous decoding, which does not need the Wyner-Ziv random binning method. Moreover, the work in \cite{10147248} reveals the deterministic-random tradeoff between \ac{snc} within the dual-functional signaling strategy by characterizing the Cram\'er-Rao bound (CRB)-communication rate region. Unfortunately, the fundamental limits of \ac{sdas} systems, whose setups significantly differ from the existing literature, remain widely unexplored \cite{9705498}. 

The information-theoretic framework for \ac{sdas} systems is depicted in Fig \ref{fig2}, where $\Rv{S}$, $\Rv{Z}$, $\Rv{X}$, $\Rv{Y}$, and $\Rv{\hat{S}}$ represent the source, sensing channel observation, \ac{snc} channels input, communication channel output, and recovery at the receiver, respectively. Essentially, the \ac{sdas} framework shares a similar methodology with conventional remote estimation (RE), remote source coding (RSC), and information bottleneck (IB) problems, by striving to balance the tradeoff between compression complexity (from $\Rv{Z}$ to $\Rv{\hat{S}}$) and the informative preservation (from $\Rv{S}$ to $\Rv{\hat{S}}$). However, we highlight the unique challenge introduced by adopting the ISAC signal $\Rv{X}$ as both the \ac{snc} channels input.

%$\bullet$ \textbf{Remote estimation:} 
\subsubsection{Remote estimation}
The RE framework involves a sensor measuring the state of a linear system and transmitting its observations to a remote estimator over a wireless fading channel \cite{6286997,8635889,gao2018optimal,7935515}. Sensing data acquisition in RE scheme is achieved through additional sensors, resulting that the observation $\Rv{Z}$ and the communication channel input $\Rv{X}$ are decoupled. In contrast, the observation $\Rv{Z}$ depends on the ISAC channel input $\Rv{X}$ due to the dual-functional signaling strategy in \ac{sdas} system.     

%$\bullet$ \textbf{Remote source coding:} 
\subsubsection{Remote source coding}
In RSC framework, the noisy observation $\Rv{Z}$ is encoded instead of the original source information $\Rv{S}$ \cite{BookBerger,623151,6915877,7464359,8636539}. Rate-distortion theory is used to quantify the tradeoff between communication rate and achievable distortion, benefiting from the Markov chain $\Rv{S} \leftrightarrow \Rv{Z} \leftrightarrow \Rv{X} \leftrightarrow \Rv{Y} \leftrightarrow \Rv{\hat{S}}$ \cite{8636539}. However, the dependency between $\Rv{Z}$ and $\Rv{X}$ in \ac{sdas} violates the Markov chain assumption, complicating the characterization of the rate-distortion relationship.      

%$\bullet$ \textbf{Information bottleneck:} 
\subsubsection{Information bottleneck}
The IB problem is essentially equivalent to the RSC problem in which the distortion metric is specified by the logarithm loss fidelity criterion \cite{9082644,e22020151}. For discussion convenience, we focus on the simplified process $\Rv{S} \leftrightarrow \Rv{Z} \leftrightarrow \Rv{\hat{S}}$. The IB problem may be expressed by
\begin{equation}
\mathop { \text{max} } \kern 2pt I(\Rv{S};\Rv{\hat{S}}) \kern 5pt  \text{subject to} \kern 5pt I(\Rv{Z};\Rv{\hat{S}}) \le \beta, \notag
\end{equation} 
where $I(\cdot,\cdot)$ denotes the mutual information (MI) and $\beta$ is the preset positive Lagrange multiplier that balances the tradeoff between compression and relevance. In contrast, when incorporating the ISAC signal $\Rv{X}$ in \ac{sdas} system, all variables in the IB problem become coupled. This leads to the following reformulation      
\begin{equation}
\mathop { \text{max} } \kern 2pt I(\Rv{S};\Rv{\hat{S}}(\Rv{X})) \kern 5pt  \text{subject to} \kern 5pt I(\Rv{Z}(\Rv{X});\Rv{\hat{S}}(\Rv{X})) \le I(\Rv{X};\Rv{Y}), \notag
\end{equation} 
where the tradeoff parameter $\beta$ is replaced by the communication channel capacity $I(\Rv{X};\Rv{Y})$ which also depends on $\Rv{X}$. Evidently, the introduced ISAC signal significantly complicates the IB problem within the context of \ac{sdas} system.

\subsection{Our contributions} 
Considering the aforementioned unique challenges, our aim is to establish the information-theoretic framework for the \ac{sdas} systems, characterize its fundamental limits, and design dedicated ISAC waveform for \ac{sdas} transmission. Compared to the conference version \cite{dong2024fundamental}, which only discussed the information-theoretic framework, we further elaborate on the optimal distribution of channel input and propose a novel method for Gaussian waveform design in MIMO \ac{sdas} systems. The main contributions are summarized as follows. 

\begin{itemize}
	\item First, we establish the information-theoretic framework for \ac{sdas} systems to illustrate the relationship between achievable distortion, coding rate, and communication channel capacity. We develop a modified source-channel separation theorem (MSST) specific to the cases of separable distortion metric for \ac{snc} processes. Compared to the existing works \cite{10845869,10233679}, we provide a rigorous proof of the operational meaning for the MSST.

	\item Second, we develop a optimal distribution design for the ISAC signal (i.e., \ac{snc} channel input), where the \ac{sdas} distortion is minimized while adhering to the MSST and resource budget constraints. To cope with the functional optimization problem, we conceive a two-step Blahut-Arimoto (BA)-based optimal search algorithm in an effort to tackle the challenges of lacking explicit expressions for the rate-distortion bound and channel capacity.    
	
	\item Third, we propose an optimal waveform design scheme for Gaussian signaling in MIMO \ac{sdas} systems. Under the Gaussian assumption, the explicit expressions of the rate-distortion function and channel capacity are derived. Thus, the optimal waveform design is modeled by a covariance matrix optimization problem solved by using an successive convex approximation (SCA)-based algorithm.

	\item Finally, we provide numerical simulation results to show the effectiveness of the proposed algorithms, and demonstrate the performance tradeoff between \ac{snc} processes in the \ac{sdas} systems. 
\end{itemize}

This paper is structured as follows. We commence with establishing the information-theoretic framework in Section \ref{FrameworkCAS}, including the definitions of distortions, rate-distortion function, and constrained channel capacity. In Section \ref{SMSST}, we prove the achievability and converse of the MSST. Then, we formulate the optimization problem for ISAC channel input design and develop a two-step BA-based search algorithm in Section \ref{Pformulation}. In Section \ref{WaveformDesign}, we present the Gaussian ISAC waveform design method for the MIMO \ac{sdas} systems. Finally, we provide the simulation results and conclude this paper in Section \ref{simulationR} and \ref{Conclude}, respectively.
 
\textbf{Notations:} The uppercase normal letter $\Rv{A}$, lowercase italic letter $a$, and fraktur letter $\mathcal{A}$ denote a random variable, its realization and a set, respectively. $\Prob{A}{a}$ represent a probability distribution function and $\Claw{A|B}{a|b}$ specifies to channel transition probability. Uppercase and lowercase bold letters $\mathbf{A}$ and $\mathbf{a}$ denote the matrix and column vector. $(\cdot)^T$, $(\cdot)^*$, and $(\cdot)^H$ represent the transpose, conjugate, and complex conjugate transpose operations, respectively. $\mathept{\cdot}$ is the statistical expectation, and $\tr{\cdot}$ is the trace of a matrix.

\section{Information-theoretic framework of \ac{sdas}} \label{FrameworkCAS}

\subsection{Sensing and Communication Processes}

As shown in Fig. \ref{fig2}, the random variables of the target's state $\Rv{S}$, the sensing observation $\Rv{Z}$, the ISAC signal $\Rv{X}$, and the communication channel output $\Rv{Y}$ take values in the sets $\mathcal{S}$, $\mathcal{Z}$, $\mathcal{X}$, and $\mathcal{Y}$, respectively. Here, the state sequence $\{\Rv{S}_i\}_{i \ge 1}$ is \ac{iid} subject to a prior distribution $\Prob{S}{s}$. In general, the \ac{sdas} system may be described as the following \ac{snc} processes.

$\bullet$ \textbf{Sensing Process}: The sensing channel output, $\Rv{Z}_i$, at a given time $i$ is generated based on the sensing channel law $\Claw{Z|XS}{\cdot|x_i,s_i}$ given the $i$th channel input $\Rv{X}_i=x_i$ and the state realization $\Rv{S}_i=s_i$. We assume that the observation $\Rv{Z}_i$ is independent of past inputs, outputs and state signals. 

$\bullet$ \textbf{Communication Process}: The ISAC signal $\Rv{X}_i$ is obtained by encoding the observations $\Rv{Z}^{i-1}$. The decoder receives the communication channel output $\Rv{Y}_i$ in accordance with channel law $\Claw{Y|X}{y|x}$ and maps $\mathcal{Y}^n$ to $\hat{\mathcal{S}}^n$, where $\hat{\mathcal{S}}$ denotes the reconstruction alphabet. 

By denoting $R$ as the bit rate, a $(2^{nR}, n)$ coding scheme for the \ac{sdas} systems consists of 

1) A message set (quantized observations) $\mathcal{Z}^n = [1:2^{nR}]$;

2) A sequence of encoding functions $\phi_i$: $\mathcal{Z}^{i-1} \to \mathcal{X}$;

3) A decoder $\psi$: $\mathcal{Y}^n \to \hat{\mathcal{S}}^n$. 

The performance of \ac{sdas} systems can be evaluated by the distortion between the ground truth and its reconstruction at the receiver. We term this distortion as the \emph{\ac{sdas} distortion}, which can be defined by
\begin{equation}
	\Delta^{(n)}:= \mathept{d(\Rv{S}^n,\hat{\Rv{S}}^n)} = \frac{1}{n}\sum_{i=1}^n \mathept{d(\Rv{S}_i,\hat{\Rv{S}}_i)},
\end{equation}  
where $d(\cdot, \cdot)$ is the distortion function bounded by $d_\text{max}$.

In practical systems, the \ac{snc} channel input $\Rv{X}$ may be restricted by limited system resources. Let us define the cost-function $b(x):\mathcal{X} \to \mathbb{R}^+$ as the channel cost. Thus, a rate-distortion-cost tuple $(R,D,B)$ is said achievable if there exists a sequence of $(2^{nR},n)$ codes that simultaneously satisfy \cite{9785593}
\begin{equation}
\varlimsup_{n \to \infty} \Delta^{(n)} \le D, \kern 5pt \varlimsup_{n \to \infty} \mathbb{E}[b(\Rv{X}^n)] \le B,
\end{equation} 
where $D$ and $B$ represent the \ac{sdas} distortion and resource budget, respectively. We define the capacity-distortion-cost function $C_\text{\ac{sdas}}(D,B)$ as the infimum of rate $R$ such that the tuple $(R,D,B)$ is achievable. As analyzed in Section \ref{relatework}, compared to the existing works \cite{8437621,9785593}, it is difficult to characterize such a capacity-distortion-cost curve due to the strong coupling between \ac{snc} process through the ISAC signal. In this paper, we characterize $C_\text{\ac{sdas}}(D,B)$ within an estimate and compress (EC) strategy which achieves an optimal tradeoff between rate and distortion \cite{8744500}, and leave the general strategy for future research.

\subsection{The Estimate and Compress Strategy }
In estimate and compress strategy, the encoder transmits estimate $\Rv{\tilde{S}}$ which is extracted from observation $\Rv{Z}$, instead of transmitting $\Rv{Z}$ itself. The coding scheme may be recast by 
	 
1) A state parameter estimator $h$: $\mathcal{X}^n \times \mathcal{Z}^n \to \tilde{\mathcal{S}}^n$, where $\tilde{\mathcal{S}}$ denote the estimate alphabet;

2) A message set (quantized estimations) $\tilde{\mathcal{S}}^n = [1:2^{nR}]$;
		
3) A encoder function $\phi$: $\tilde{\mathcal{S}}^n \to \mathcal{X}^n$.

4) A decoder $\psi$: $\mathcal{Y}^n \to \hat{\mathcal{S}}^n$.

%Thus, we may define the estimation distortion in the sensing process by
%\begin{equation}
%	\Delta_s^{(n)}:= \mathept{d(\Rv{S}^n,\tilde{\Rv{S}}^n)}=\frac{1}{n}\sum_{i=1}^n \mathept{d(\Rv{S}_i,\tilde{\Rv{S}}_i)},
%\end{equation} 
%and the recovery distortion in the communication process by
%\begin{equation}
%	\Delta_c^{(n)}:= \mathept{d(\tilde{\Rv{S}}^n,\hat{\Rv{S}}^n)} =\frac{1}{n}\sum_{i=1}^n \mathept{d(\tilde{\Rv{S}}_i,\hat{\Rv{S}}_i)}.
%\end{equation} 

Although the estimator and encoder are defined in a block form for generality, the following Lemma 1 \cite{9785593} show that the optimal estimator $h$ can be achieved through a symbol-by-symbol estimation.

\textbf{\underline{\textit{Lemma 1}}} \cite{9785593}: By recalling that $\Rv{S} \leftrightarrow \Rv{XZ} \leftrightarrow \Rv{\tilde{S}}$ forms a Markov chain, the sensing distortion $\Delta_s^{(n)}$ is minimized by the deterministic estimator
\begin{equation}\label{opest}
	h^\star(x^n,z^n):= (\tilde{s}^\star(x_1,z_1),\tilde{s}^\star(x_2,z_2),\cdots,\tilde{s}^\star(x_n,z_n)),
\end{equation}
where
\begin{equation} \label{escost}
	\tilde{s}^\star(x,z):= \arg \min_{s'\in \mathcal{\tilde{S}}} \sum_{s \in \mathcal{S}}\Claw{S|XZ}{s|x,z}d(s,s'),
\end{equation}
with posterior transition probability being
\begin{equation}
\Claw{S|XZ}{s|x,z} = \frac{\Prob{S}{s} \Claw{Z|XS}{z|x,s}}{\sum_{s' \in \mathcal{S}}\Prob{S}{s'} \Claw{Z|XS}{z|x,s'}} , \notag
\end{equation}
which is independent to the choice of encoder and decoder. The detailed proof can be found in \cite{9785593}. By applying the optimal estimator \eqref{escost}, the estimate $\Rv{\tilde{S}}_i$ only depends on $\Rv{X}_i$ and $\Rv{Z}_i$ but independent of past inputs and outputs. Thus, the optimal encoder is reduced to a symbol-wise encoding process. 

By applying the estimate and compress strategy, the information flow of the \ac{sdas} can be divided into two parts: 1) Obtain the estimate $\Rv{\tilde{S}}_i$ from $\Rv{Z}_i$ in sensing process; 2) Transmit the estimate $\Rv{\tilde{S}}_i$ and recover $\Rv{\hat{S}}_i$ from $\Rv{Y}_i$ in communication process. Let us define the \ac{snc} distortions as the expectations of the estimate error and communication error, i.e.,
\begin{equation}
D_s =\mathept{d(\Rv{S},\Rv{\tilde{S}})}, \kern 5pt D_c = \mathept{d(\tilde{\Rv{S}},\Rv{\hat{S}})}.
\end{equation} 	      
Next, we will show how \ac{snc} distortions can be achievable. 	
	
Let us define the estimate error function in sensing process \ac{wrt} the channel input realization $\Rv{X}=x$ as
\begin{equation}\label{exd}
e(x) = \mathept{d(\Rv{S},\tilde{s}^\star(\Rv{X},\Rv{Z}))|\Rv{X}=x}.
\end{equation} 
Thus, the sensing distortion $D_s$ is achievable when $\mathept{e(\Rv{X})} \le D_s$ holds. Nevertheless, communication distortion $D_c$ strictly depends on channel capacity which in turn relies on the channel input $\Rv{X}$. Thus, the strong coupling property, stemming from the shared channel input $\Rv{X}$, leads to an inherent performance tradeoff between the \ac{snc} distortions. To quantify such relationship, we define the communication channel capacity for a given sensing distortion as follows.  

\underline{\textit{\textbf{Definition 1}}}: The communication channel capacity constrained by sensing distortion $D_s$ and resource cost $B$ can be defined by  
\begin{equation}\label{CapTrade}
	C_\text{IT}(D_s, B) = \opmax{\Prob{X}{x} \in \mathcal{P}_{D_s} \cap \mathcal{P}_B} I(\Rv{X};\Rv{Y}),
\end{equation}    
where $\mathcal{P}_{D_s}$ is the sensing feasible probability set whose element satisfies
\begin{equation}
\mathcal{P}_{D_s}= \left\{\Prob{X}{x}\big|\mathept{e(\Rv{X})} \le D_s \right\}, 
\end{equation}  
and $\mathcal{P}_B$ is the resource feasible probability set described by
\begin{equation}
\mathcal{P}_B=\left\{\Prob{X}{x}\big|\mathept{b(\Rv{X})} \le B \right\}.
\end{equation}    

The capacity restricts the maximum achievable rate $R$ of the communication channel. It is well-known that the relationship between rate $R$ and distortion $D_c$ can be characterized by the rate-distortion theory. Before presenting the rate-distortion function, we first show the communication source $\{\tilde{\Rv{S}}_i\}_{i \ge 1}$ is \ac{iid} with the distribution    
\begin{equation}\label{PtildeS}
	\Prob{\tilde{\Rv{S}}}{\tilde{\Rv{s}}} = \sum_{z \in \mathcal{Z}} \sum_{x \in \mathcal{X}} \sum_{s \in \mathcal{S}} \Claw{Z|XS}{z|x,s}\Prob{S}{s}\Prob{X}{x}\mathbb{I} \{\tilde{s}^\star(x,z)=\tilde{s}\},
\end{equation}	
with $\mathbb{I}(\cdot)$ being the indicator function. 

Note that the sequence $\{\Rv{X}_i\}_{i \ge 1}$ is \ac{iid} with distribution $\Prob{X}{x}$ for achieving the capacity-distortion-cost region. Thus, the observation $\{\Rv{Z}_i\}_{i \ge 1}$ is also \ac{iid} due to the \ac{iid} target's state $\{\Rv{S}_i\}_{i \ge 1}$. By adopting the symbol-by-symbol optimal estimator \eqref{escost}, we can immediately obtain \eqref{PtildeS}. Therefore, for \ac{iid} source $\Rv{\tilde{S}}$, the rate-distortion function for the communication process can be defined as follows.  
      
%\begin{equation}
%\Prob{Z}{z}=\sum_{s \in \mathcal{S}} \sum_{x \in \mathcal{X}} \Claw{Z|XS}{z|x,s}\Prob{S}{s}\Prob{X}{x}.
%\end{equation}   

\underline{\textit{\textbf{Definition 2}}}: The information-theoretic rate distortion function can be defined by 
\begin{equation}\label{srd}
	R_\text{IT}(D_c) = \opmin{\Prob{\hat{S}|\tilde{S}}{\hat{s}|\tilde{s}}: \mathept{d(\Rv{\tilde{S}},\Rv{\hat{S}})} \le D_c} I(\Rv{\tilde{S}};\Rv{\hat{S}}),
\end{equation}
where the minimization is over all conditional distributions $\Prob{\Rv{\hat{S}}|\Rv{\tilde{S}}}{\hat{s}|\tilde{s}}$ for which the joint distribution $\Prob{\tilde{S}\hat{S}}{\tilde{s},\hat{s}} =\Prob{\tilde{\Rv{S}}}{\tilde{\Rv{s}}} \Prob{\Rv{\hat{S}}|\Rv{\tilde{S}}}{\hat{s}|\tilde{s}}$ satisfies the expected distortion constraint.

According to the conventional source-channel separation theorem with distortion in lossy data transmission \cite[Theorem 10.4.1]{thomas2006elements}, we have the conclusion that the communication distortion $D_c$ is achievable if and only if $R_\text{IT}(D_c) \le C_\text{IT}(D_s, B)$. Building upon this, the remaining question is how to define the \ac{sdas} distortion in terms of the \ac{snc} distortions.

\section{Modified Source-Channel Separation Theorem} \label{SMSST}

In this section, we develop the modified source-channel separation theorem tailored for the \ac{sdas} systems, aiming to elucidate the relationships between the \ac{sdas} distortion, coding rate and channel capacity. In the MSST, we restrict the distortion metric satisfying the following separable condition.     

\underline{\textit{\textbf{Definition 3}}}: The distortion metric $d(\cdot, \cdot)$ is separable for the \ac{snc} processes, if the following equality holds,
\begin{equation}\label{separablec}
	\mathept{d(\Rv{S},\Rv{\hat{S}})} = \mathept{d(\Rv{S},\Rv{\tilde{S}})} + \mathept{d(\tilde{\Rv{S}},\Rv{\hat{S}})},
\end{equation}      
implying that the \ac{sdas} distortion $D$ is equal to $D_s+D_c$.
 
Intuitively, the separable condition does not necessarily hold in most scenarios. Fortunately, we demonstrate that the \ac{mse} or quadratic distortion, a widely used distortion metric in parameter estimation, satisfying 
\begin{equation} \label{qudraticmtric}
 \mathept{\normquad{\Rv{S}-\hat{\Rv{S}}}}  \mathop = \limits^{(a)} \mathbb{E}\left[\left\|\mathsf{S}- \tilde{\mathsf{S}}\right\|_2^2 \right]+\mathbb{E}\left[\left\|\tilde{\mathsf{S}} - \hat{\mathsf{S}}\right\|_2^2 \right],
\end{equation}     
where $(a)$ holds from the properties of the conditional expectation \cite[Appendix A]{8636539} with $\mathsf{\tilde{S}}=\mathbb{E}\left[ \mathsf{S}|\mathsf{Z}, \mathsf{X} \right]$ being the optimal estimator in \eqref{escost}. Now, we are ready to propose MSST \footnote{We highlight that the \eqref{escost} is exactly the minimum MSE (MMSE) estimator. An arbitrary non-optimal estimator except MMSE, which cannot apply the conditional expectation properties, may not satisfy the separable condition.}.

\underline{\textit{\textbf{Theorem 1}}}: The \ac{sdas} distortion $D=D_s+D_c$ is achievable within a separable distortion metric, if and only if 
\begin{equation}\label{theorem}
R_\text{IT}(D_c) \le C_\text{IT}(D_s, B),
\end{equation}
where $C_\text{IT}(D_s, B)$ and $R_\text{IT}(D_c)$ are constrained channel capacity \eqref{CapTrade} and rate-distortion function \eqref{srd}, respectively. 

The above MSST is proposed mathematically within the framework of information theory. Next, we elucidate the operational meaning of the MSST by proving that there must exist a practical coding scheme satisfying it.

\subsection{Converse}
We start with a converse to show that any achievable coding scheme must satisfy \eqref{theorem}. Consider a $(2^{nR},n)$ coding scheme defined by the encoding and decoding functions $\phi$ and $\psi$. Let $\Rv{\tilde{S}}^n = h^\star(\Rv{X}^n,\Rv{Z}^n)$ be the estimate sequence as given in \eqref{opest} and $\Rv{\hat{S}}^n = \psi(\phi(\Rv{\tilde{S}}^n))$ be the reconstruction sequence. 

Let us focus on the communication process $\Rv{\tilde{S}} \leftrightarrow \Rv{X} \leftrightarrow \Rv{Y} \leftrightarrow \Rv{\hat{S}}$. By recalling from the proof of the converse in lossy source coding, we have
\begin{equation}\label{r1}
	\begin{aligned}
		R &\ge \frac{1}{n}\sum_{i=1}^{n} I(\Rv{\tilde{S}}_i;\Rv{\hat{S}}_i) \mathop{\ge} \limits^{(a)}  \frac{1}{n}\sum_{i=1}^{n} R_\text{IT} \left( \mathept{d(\Rv{\tilde{S}}_i,\Rv{\hat{S}}_i)} \right) \\
		& \mathop{\ge} \limits^{(b)}  R_\text{IT} \Big(\frac{1}{n} \sum_{i=1}^{n} \mathept{d(\Rv{\tilde{S}}_i,\Rv{\hat{S}}_i)} \Big) \mathop{\ge} \limits^{(c)} R_\text{IT}(D_c),
	\end{aligned}
\end{equation}
where $(a)$ follows the Definition 2 that $R_\text{IT}$ is the minimum required MI, $(b)$ and $(c)$ are due to the convexity and non-increasing properties of the rate distortion function. 

On the other hand, by recalling from the proof of the converse in channel coding, we have
\begin{equation}\label{r2}
	\begin{aligned}
		R &\le \frac{1}{n}\sum_{i=1}^{n} I(\Rv{X}_i;\Rv{Y}_i) \\
		& \mathop{\le} \limits^{(d)}  \frac{1}{n}\sum_{i=1}^{n} C_\text{IT} \Big( \sum_{x\in\mathcal{X}}\Prob{X_i}{x}c(x), \sum_{x\in\mathcal{X}} \Prob{X_i}{x}b(x)\Big) \\
		& \mathop{\le} \limits^{(e)}  C_\text{IT} \Big(\frac{1}{n}\sum_{i=1}^{n}\sum_{x\in\mathcal{X}}\Prob{X_i}{x}c(x), \frac{1}{n}\sum_{i=1}^{n}\sum_{x\in\mathcal{X}}\Prob{X_i}{x}b(x)\Big) \\
		& \mathop{\le} \limits^{(f)} C_\text{IT}(D_s,B),
	\end{aligned}
\end{equation}  
where $(d)$ follows the Definition 1 that the channel capacity is the maximum MI, $(e)$ and $(f)$ are due to the concavity and non-decreasing properties of the capacity constrained by estimation and resource costs \cite{9785593}. By combing the inequalities \eqref{r1}, \eqref{r2} and the data processing inequality $I(\Rv{\tilde{S}};\Rv{\hat{S}}) \le I(\Rv{X};\Rv{Y})$, we complete the proof of the converse.  

\subsection{Achievability}
Let $\Rv{S}^n$ be drawn \ac{iid} $\sim \Prob{S}{s}$, we will show that there exists a coding scheme for a sufficiently large $n$ and rate $R$, the distortion $\Delta^{n}$ can be achieved by $D$ if \eqref{theorem} holds. The core idea follows the famous \emph{random coding argument} and \emph{source channel separation theorem with distortion}.  

1) \textit{Codebook Generation}: In source coding with rate distortion code, randomly generate a codebook $\mathcal{C}_s$ consisting of $2^{nR}$ sequences $\Rv{\hat{S}}^n$ which is drawn \ac{iid} $\sim \Prob{\hat{S}}{\hat{s}}$. The probability distribution is calculated by $\Prob{\hat{S}}{\hat{s}} = \sum_{\tilde{s} }\Prob{\tilde{S}}{\tilde{s}}\Claw{\hat{S}|\tilde{S}}{\hat{s}|\tilde{s}}$, where $\Prob{\tilde{S}}{\tilde{s}}$ is defined in \eqref{PtildeS} and $\Claw{\hat{S}|\tilde{S}}{\hat{s}|\tilde{s}}$ achieves the equality in \eqref{srd}. In channel coding, randomly generate a codebook $\mathcal{C}_c$ consisting of $2^{nR}$ sequences $\Rv{X}^n$ which is drawn \ac{iid} $\sim \Prob{X}{x}$. The $\Prob{X}{x}$ is chosen by satisfying the constrained capacity with estimation- and resource-cost in \eqref{CapTrade}. Index the codeword $\Rv{\tilde{S}}^n$ and $\Rv{X}^n$ by $w \in \{1,2,\cdots, 2^{nR}\}$. 

2) \textit{Encoding}: Encode the $\Rv{\tilde{S}}^n$ by $w$ such that
\begin{equation}
	(\Rv{\tilde{S}}^n, \Rv{\hat{S}}^n(w)) \in \mathcal{T}^{(n)}_{d,\epsilon_s} (\Prob{\tilde{S}\hat{S}}{\tilde{s},\hat{s}}),
\end{equation}
where $\mathcal{T}^{(n)}_{d,\epsilon_s} (\Prob{\tilde{S}\hat{S}}{\tilde{s},\hat{s}})$ represents the distortion typical set \cite{thomas2006elements} with joint probability distribution $\Prob{\tilde{S}\hat{S}}{\tilde{s},\hat{s}} = \Prob{\tilde{S}}{\tilde{s}}\Claw{\hat{S}|\tilde{S}}{\hat{s}|\tilde{s}}$. To send the message $w$, the encoder transmits $x^n(w)$. 

3) \textit{Decoding}: The decoder observes the communication channel output $\Rv{Y}^n=y^n$ and look for the index $\hat{w}$ such that 
\begin{equation}
	(x^n(\hat{w}), y^n) \in \mathcal{T}^{(n)}_{\epsilon_c}(\Prob{XY}{x,y}),
\end{equation} 
where $\mathcal{T}^{(n)}_{\epsilon_c}(\Prob{XY}{x,y})$ represents the typical set with joint probability distribution $\Prob{XY}{x,y}=\Prob{X}{x}\Claw{Y|X}{y|x}$.  If there exists such $\hat{w}$, it declares $\Rv{\hat{S}}^n = \hat{s}^n(\hat{w})$. Otherwise, it declares an error. 

4) \textit{Estimation}: The encoder observes the channel output $\Rv{Z}^n = z^n$, and computes the estimate sequence with the knowledge of channel input $x^n$ by using the estimator $\tilde{s}^n = h^\star(x^n,z^n)$ given in \eqref{opest}. 

5) \textit{Distortion Analysis}: We start by analyzing the expected communication distortion (averaged over the random codebooks, state and channel noise). In lossy source coding, for a fixed codebook $\mathcal{C}_s$ and choice of $\epsilon_s > 0$, the sequence $\tilde{s}^n \in \tilde{\mathcal{S}}^n$ can be divided into two categories:

$\bullet$ $(\tilde{s}^n, \hat{s}^n(w)) \in \mathcal{T}^{(n)}_{d,\epsilon_s}$, we have $d(\tilde{s}^n, \hat{s}^n(w)) < D_c + \epsilon_s$; 

$\bullet$ $(\tilde{s}^n, \hat{s}^n(w)) \notin \mathcal{T}^{(n)}_{d,\epsilon_s}$, we denote $P_{e_s}$ as the total probability of these sequences. Thus, these sequence contribute at most $P_{e_s}d_{\max}$ to the expected distortion since the distortion for any individual sequence is bounded by $d_{\max}$.    

According to the achievability of lossy source coding \cite[Theorem 10.2.1]{thomas2006elements}, we have $P_{e_s}$ tends to zero for sufficiently large $n$ whenever $R \ge R_\text{IT}(D_c)$. 

In channel coding, the decoder declares an error when the following events occur:

$\bullet$  $(x^n (w), y^n) \notin \mathcal{T}^{(n)}_{\epsilon_c}$; 

$\bullet$  $(x^n (w'), y^n) \in  \mathcal{T}^{(n)}_{\epsilon_c}$, for some $w' \ne w$, we denote $P_{e_c}$ as the probability of the error occurred in decoder. The error decoding contribute at most $P_{e_c}d_{\max}$ to the expected distortion. Similarly, we have $P_{e_c} \to 0$ for $n \to \infty$ whenever $R \le C_\text{IT}(D_s, B)$ according to channel coding theorem \cite{thomas2006elements}. 

On the other hand, the expected estimation distortion can be upper bounded by
\begin{equation}
	\begin{aligned}
		&\Delta_s^{(n)} = \frac{1}{n}\sum_{i=1}^n \mathept{d(\Rv{S}_i,\Rv{\tilde{S}}_i)|\Rv{\hat{W}} \ne w} \text{Pr}(\Rv{\hat{W}} \ne w) \\
		& \kern 30pt + \frac{1}{n}\sum_{i=1}^n \mathept{d(\Rv{S}_i,\Rv{\tilde{S}}_i)|\Rv{\hat{W}} = w} \text{Pr}(\Rv{\hat{W}} = w) \\
		& \le P_{e_c}d_\text{max} + \frac{1}{n}\sum_{i=1}^n \mathept{d(\Rv{S}_i,\Rv{\tilde{S}}_i)|\Rv{\hat{W}} = w} (1-P_{e_c}). 
	\end{aligned}
\end{equation}

Note that $(s^n, x^n(w), \tilde{s}^n) \in \mathcal{T}^{(n)}_{\epsilon_e} (\Prob{SX\tilde{S}}{s,x,\tilde{s}})$ where $\Prob{SX\tilde{S}}{s,x,\tilde{s}}$ denotes the joint marginal distribution of $\Prob{SXZ\tilde{S}}{s,x,z,\tilde{s}} = \Prob{S}{s}\Prob{X}{x}\Claw{Z|SX}{z|s,x} \mathbb{I} \{\tilde{s}=\tilde{s}^\star(x,z)\}$, we have 
\begin{equation}
	\varlimsup_{n \to \infty} \frac{1}{n}\sum_{i=1}^n \mathept{d(\Rv{S}_i,\Rv{\tilde{S}}_i)|\Rv{\hat{W}} = w} \le (1+\epsilon_e) \mathept{d(\Rv{S},\Rv{\tilde{S}})},
\end{equation}    
according to the typical average lemma \cite{9785593}. In summary, the \ac{sdas} distortion can be attained by
\begin{equation}
	\begin{aligned}
		\Delta^{(n)} & \mathop{=} \limits^{(a)}  \Delta_s^{(n)} + \Delta_c^{(n)} \\
		& \mathop{\le} \limits^{(b)} D_c + (P_{e_s}+ 2P_{e_c})d_{\max} + (1+\epsilon_e)(1-P_{e_c})D_s, 
	\end{aligned}
\end{equation} 
where $(a)$ follows the condition of the separable distortion metric in MSST and we omit the terms containing the product of $P_{e_s}$ and $P_{e_c}$ in step $(b)$. Consequently, taking $n \to \infty$ and $P_{e_s}, P_{e_c}, \epsilon_c,\epsilon_e,\epsilon_s \to 0$, we can conclude that the expected SAS distortion (averaged over the random codebooks, state and channel noise) tends to be $D \to D_c+D_s$ whenever $ R_\text{IT}(D_c) \le R \le C_\text{IT}(D_s, B)$.  

This completes the proof of the proposed MSST. $\hfill\blacksquare$ 

{\textbf{Remark}:} Intuitively, the sensing process determines the accuracy of the state information acquisition at the encoder, whereas the communication process governs the quantity of information transmitted to the decoder. Overemphasis on either process can result in substantial performance degradation of the \ac{sdas} system. \textit{Theorem 1} and its proof quantitatively show the condition (i.e., formula \eqref{theorem}) required to achieve the \ac{sdas} distortion $D$. Although the separable condition in \eqref{separablec} is to some extend strong, \textit{Theorem 1} is applicable in most parameter estimation scenarios which adopts MMSE estimator.   

Additionally, \textit{Theorem 1} provides us a powerful tool to investigate the optimal ISAC channel input $\Rv{X}$. Specifically, achieving optimal sensing performance alone may not ensure the accurate reconstruction at the decoder due to limitations in the communication channel capacity. Conversely, excessive communication capacity may be wasted if the sensory data lacks sufficient accuracy. This underscores the need to minimize the \ac{sdas} distortion by optimizing the distribution (Section \ref{Pformulation}) and the waveform (Section \ref{WaveformDesign}) in next sections.

\section{Channel Input Distribution Optimization} \label{Pformulation}

In this section, we explore the optimal input distribution for the \ac{sdas} system under scalar channels, aiming to minimize the \ac{sdas} distortion under the MMST constraint. Hereinafter, we adopt \ac{mse} as the distortion metric to guarantee the MSST.     

\subsection{Problem Formulation}

The ISAC channel input design can be formulated by the following optimization problem \footnote{Here, we use the symbol $I(D_s,B)$ instead of $C(D_s,B)$ to emphasize that the optimal distribution does not necessarily achieve the channel capacity.}
\begin{equation} \label{p0}
\mathcal{P}_0  \left\{ \kern 2pt
\begin{aligned}
\mathop { \text{min} }\limits_{P_\mathsf{X}(x)} \kern 5pt & D = D_c + D_s   \\
\text{subject to} \kern 5pt & R(D_c) \le I(D_s, B),
\end{aligned} \right.
\end{equation} 
where the definitions of $R(D_c)$ and $I(D_s, B)$ are given in \eqref{srd} and \eqref{CapTrade}, respectively. Here, we drop the subscript ``IT'' to emphasize the operational meanings is justified by the MSST from the perspective of coding scheme. Furthermore, we use the mutual information symbol $I(D_s, B)$ instead of $C(D_s,B)$ to emphasize that the optimal distribution of $\mathcal{P}_0$ does not necessarily achieve the channel capacity. This modification is just to keep the math rigorous, but it does not change the expression. By substituting the associated expressions into \eqref{p0}, problem $\mathcal{P}_0$ can be reformulated by
\begin{equation}
	\mathcal{P}_1  \left\{ \kern 2pt
	\begin{aligned}
		\mathop { \text{min} }\limits_{P_\mathsf{X}(x)} \kern 5pt & \mathept{d(\Rv{S},\tilde{\Rv{S}})} + \mathept{d(\Rv{\tilde{S}},\hat{\Rv{S}})}    \\
		\text{subject to} \kern 5pt & I(\mathsf{\tilde{S};\hat{S}}) \le I(\Rv{X};\Rv{Y}), \\
						  \kern 5pt &  \mathept{b(\Rv{X})} \le B.
	\end{aligned} \right.
\end{equation}
It is challenging to directly solve the functional optimization problem $\mathcal{P}_1$ due to the difficulty in obtaining explicit expressions of distortions and MI for arbitrary distributions. Typically, the BA algorithm can be employed to solve problem $\mathcal{P}_1$ numerically. Inspired by this idea, we expand the expectation operation and the MI \ac{wrt} the variable $\Prob{X}{x}$ as follows.

$\bullet$ \textit{Sensing distortion:}
\begin{equation}
\begin{aligned}
\mathept{d(\Rv{S},\Rv{\tilde{S}})} &= \mathept{\mathept{d(\mathsf{S},\tilde{s}(\mathsf{X},\mathsf{Z}))|\mathsf{X},\mathsf{Z}}} \\
&= \sum_{x,z} P_{\mathsf{X}\mathsf{Z}}(x,z) \sum_s Q_{\mathsf{S}|\mathsf{X}\mathsf{Z}}(s|x,z) d(s,\tilde{s}(x,z)) \\
& \triangleq \sum_x P_\mathsf{X}(x) e(x),
\end{aligned}
\end{equation}
where the cost $e(x)$ defined in \eqref{exd} can be expressed by  
\begin{equation} \label{sensingcost}
e(x)=\sum_z Q_{\mathsf{Z}|\mathsf{X}}(z|x) \sum_s Q_{\mathsf{S}|\mathsf{X}\mathsf{Z}}(s|x,z) d(s,\tilde{s}(x,z)). 
\end{equation}

$\bullet$ \textit{Channel MI:}
\begin{equation}
\begin{aligned}
I(\mathsf{X;Y}) =  \sum_{x,y} P_\mathsf{X}(x) Q_\mathsf{Y|X}(y|x) \log \frac{Q_\mathsf{X|Y}(x|y)}{P_\mathsf{X}(x)}. \\  	
\end{aligned}
\end{equation}

$\bullet$ \textit{Resource budget:}
\begin{equation}
\mathept{b(\Rv{X})} = \sum_x P_\mathsf{X}(x)b(x).
\end{equation}

$\bullet$ \textit{Communication distortion:}
\begin{equation}\label{28a}
\begin{aligned}
\mathbb{E}[d(\mathsf{\tilde{S},\hat{S}})] = \sum_{\tilde{s}} P_\mathsf{\tilde{S}}(\tilde{s}) \sum_{\hat{s}} Q_\mathsf{\hat{S}|\tilde{S}}(\hat{s}|\tilde{s})d(\tilde{s},\hat{s}). 
\end{aligned}
\end{equation}    

$\bullet$ \textit{Rate distortion function:}
\begin{equation}\label{29a}
\begin{aligned}
I(\mathsf{\tilde{S};\hat{S}}) = \sum_{\tilde{s},\hat{s}} P_\mathsf{\tilde{S}}(\tilde{s}) Q_\mathsf{\hat{S}|\tilde{S}}(\hat{s}|\tilde{s}) \log \frac{Q_\mathsf{\tilde{S}|\hat{S}}(\tilde{s}|\hat{s})}{P_\mathsf{\tilde{S}}(\tilde{s})}. 
\end{aligned}
\end{equation}

Here, we emphasize the unique challenge in solving problem $\mathcal{P}_1$ compared to the conventional BA algorithm. The primary approach involves constructing the Lagrangian function \ac{wrt} the variable and obtaining the optimal solution based on the first-order necessary condition. However, the probability distribution $\Prob{\tilde{S}}{\tilde{s}}$ in \eqref{28a} and \eqref{29a} is indeed a function of the variable $\Prob{X}{x}$ as described in \eqref{PtildeS}. Consequently, deriving a closed-form expression of $\Prob{X}{x}$ from first-order necessary condition is challenging, which imposes a significant obstacle to solving problem $\mathcal{P}_1$. To address this issue, we propose a two-step BA-based optimal search method to seek for a sub-optimal solution in the following subsection.      

\subsection{Two-step BA-based Optimal Search Algorithm}

We divide the original problem $\mathcal{P}_1$ into two sub-problems.   

1) \textit{Sub-problem 1}: We determine the constrained communication channel capacity $I(\Rv{X},\Rv{Y})$ for a given sensing distortion $D_s$ and identify the optimal distribution $P_\mathsf{X}(x)$.  

2) \textit{Sub-problem 2}: We calculate the minimum communication distortion $D_c$ for the given channel input distribution $P_\mathsf{X}(x)$ and determine the communication distortion $D_c$, thereby obtaining the \ac{sdas} distortion $D$.

By varying the preset sensing distortions, one may generate a set $\mathcal{D}$ collecting the \ac{sdas} distortions calculated through the above two steps. Thus, the minimum \ac{sdas} distortion and its corresponding distribution $P_\mathsf{X}(x)$ can be identified by the minimum value search. The detailed procedure is as follows.

$\bullet$ \textbf{Constrained communication channel Capacity.}

In this sub-problem, $I(\mathsf{X};\mathsf{Y})$ is maximized under the constraints of sensing distortion $D_s$ and resource budget $B$, which can be expressed by   

\begin{equation} \label{maxCMI1}
	\begin{aligned}
		\mathop { \text{max} }\limits_{P_\mathsf{X}(x)} \kern 10pt & I(\mathsf{X};\mathsf{Y}) \\
		\text{subject to} & \kern 5pt \sum_x P_\mathsf{X}(x) e(x) \le D_s, \kern 5pt  \sum_x P_\mathsf{X}(x)b(x) \le B.
	\end{aligned} 
\end{equation}
Inspired by \cite{8437621}, we reformulate \eqref{maxCMI1} by incorporating the sensing distortion cost as a penalty term in the objective function, i.e.,
\begin{equation} \label{maxCMI2}
	\begin{aligned}
		\mathop { \text{max} }\limits_{P_\mathsf{X}(x)} \kern 5pt & I(P_\mathsf{X}(x), Q_\mathsf{Y|X}(y|x)) - \mu \sum_x P_\mathsf{X}(x) e(x)  \\
		\text{subject to} & \kern 5pt  \sum_x P_\mathsf{X}(x)b(x) \le B,
	\end{aligned} 
\end{equation}      
where $\mu$ is the penalty factor to balance the weight of \ac{snc} performance. For a given $\mu$, the Lagrangian function of \eqref{maxCMI2} can be written by 
\begin{equation} 
\begin{aligned}
\mathcal{L}(P_\mathsf{X}(x), \lambda) & = I(P_\mathsf{X}(x), Q_\mathsf{Y|X}(y|x)) \\
&- \mu \sum_x P_\mathsf{X}(x) e(x) + \lambda\sum_x P_\mathsf{X}(x)b(x).
\end{aligned} 
\end{equation} 
By setting the derivative $\partial \mathcal{L}(P_\mathsf{X}(x), \lambda) / \partial P_\mathsf{X}(x) $ to zero and taking the fact that $ \sum_x P_\mathsf{X}(x)=1$ into account, we have   
\begin{equation} \label{updatepx}
P_\mathsf{X}(x) = \frac{e^{h(x)}}{\sum_{x'} e^{h(x')}},
\end{equation}
with
\begin{equation} \label{hx}
h(x) = \sum_y Q_\mathsf{Y|X}(y|x) \log Q_\mathsf{X|Y}(x|y)-\mu e(x)-\lambda b(x).
\end{equation}

In formula \eqref{updatepx}, the Lagrangian multiplier $\lambda$ and the posterior distribution $Q_\mathsf{X|Y}(x|y)$ are unknown. The former may be determined by a bisection search method such that $\sum_x P_\mathsf{X}(x)b(x) \to B $. For the latter, we have the optimal posterior distribution which maximizes the MI with given $P_\mathsf{X}(x)$ and $Q_\mathsf{Y|X}(y|x)$, expressed by \cite[Lemma 9.1]{yeung2008information} 
\begin{equation} \label{postupdate}
Q_\mathsf{X|Y}(x|y) =  \frac{P_\mathsf{X}(x)Q_\mathsf{Y|X}(y|x)}{\sum_{x'}P_\mathsf{X}(x')Q_\mathsf{Y|X}(y|x')}.
\end{equation}
After setting an initial distribution $\Prob{X_0}{x}$, problem \eqref{maxCMI2} can be solved by updating \eqref{postupdate} and \eqref{updatepx} iteratively. The detailed procedure is summarized in Algorithm \ref{alg1}, whose convergence can be guaranteed \cite{9785593,li2024analysis}. 

\begin{algorithm}[!t]
	\caption{Modified Iteration BA Algorithm}
	\label{alg1}
	\KwIn{$P_\mathsf{S}(s)$, $Q_\mathsf{Z|SX}(z|s,x)$ and $Q_\mathsf{Y|X}(y|x)$, $B$, $\mu$, $K$.}
	\textbf{Initialize}: Set $P_\mathsf{X_0}(x)$ as uniform distribution; \\
	\For{$k \le K$}{
		\textbf{Step 1}: Compute sensing cost $e(x)$ in \eqref{sensingcost} \;
		\textbf{Step 2}: Update posterior distribution in \eqref{postupdate} \;
		\textbf{Step 3}: Update $P_\mathsf{X}(x)$ in \eqref{updatepx} \;   
		\textbf{Step 4}: Find $\lambda^\star$ by bisection searching \;  		
	}
	\KwOut{Optimal $P_\mathsf{X}(x)$, $D_s$ and $I(\mathsf{X};\mathsf{Y})$.}
\end{algorithm}

$\bullet$ \textbf{Source distribution $\Prob{\tilde{S}}{\tilde{s}}$.}

The tradeoff between sensing distortion and constrained communication capacity can be adjusted by varying the penalty factor $\mu$. Let us denote the optimal channel input distribution, sensing distortion, and channel MI obtained by Algorithm \ref{alg1} as $P_\mathsf{X}^{(\mu)}(x)$, $D_s^{(\mu)}$, and $I^{(\mu)}(\Rv{X};\Rv{Y})$, respectively, to highlight their dependence on the factor $\mu$. 

Subsequently, given a specific channel input distribution $P_\mathsf{X}^{(\mu)}(x)$, our objective is to compute the estimate distribution $P_\mathsf{\tilde{S}}^{(\mu)}(\tilde{s})$, which is essential for determining the rate-distortion function. However, deriving the explicit expression of the estimate distribution for an arbitrary state distribution $\Prob{S}{s}$ is a challenging task. Instead, we can compute the estimate $\tilde{\Rv{S}}$ and its distribution $P_\mathsf{\tilde{S}}^{(\mu)}(\tilde{s})$ using formulas \eqref{escost} and \eqref{PtildeS}, respectively, by performing a sufficiently large number of random trials. For analytical convenience, we consider a Gaussian state distribution and a linear sensing model, which allows for explicit expressions to be derived.  

Assume that the target's state follows Gaussian distribution with $\Prob{S}{s} = \mathcal{CN}(0,\nu_s^2)$. Let us consider the following linear sensing model                
\begin{equation} 
\mathsf{Z}=\mathsf{XS}+\mathsf{N},
\end{equation} 
where $\mathsf{N} \sim \mathcal{CN}(0,1)$ represents Gaussian channel noise. In such model, the estimator \eqref{escost} is indeed an MMSE estimator for each channel realization $x$. Therefore, the estimate $\Rv{\tilde{S}}$ and the corresponding conditional probability distribution of $P_\mathsf{\tilde{S}|X}(\tilde{s}|x)$ can be expressed by  

\begin{equation} \label{SSN}
\mathsf{\tilde{S}} = \frac{x^2\nu_s^4}{(1+x^2\nu_s^2)^2}\Rv{Z},\kern 5pt P_\mathsf{\tilde{S}|X}(\tilde{s}|x) \sim \mathcal{CN}(0,\frac{x^2\nu_s^4}{1+x^2\nu_s^2}).  
\end{equation}     
Moreover, sensing cost $e(x)$ is MSE with given realization $x$, which can be written by 
\begin{equation}
e(x)= \frac{\nu_s^2}{(1+x^2\nu_s^2)}.
\end{equation}
Consequently, the source distribution can be given by    
\begin{equation} \label{GaussianPS}
P_\mathsf{\tilde{S}}^{(\mu)}(\tilde{s}) = P_\mathsf{X}^{(\mu)}(x)P_\mathsf{\tilde{S}|X}(\tilde{s}|x).
\end{equation}

\begin{algorithm}[!t]
	\caption{Communication Distortion Algorithm}
	\label{alg2}
	\KwIn{$P_\mathsf{\tilde{S}}(\tilde{s})$, $I^{(u)}(\Rv{X};\Rv{Y})$, $K$, a set of the slops $\mathcal{A}$.}
	\For{$\lambda_s \in \mathcal{A}$}{
		\textbf{Initialize}: Set $Q_\mathsf{\hat{S}|\tilde{S}}(\hat{s}|\tilde{s})$ as uniform distribution ; \\
		\For{$k \le K$}{
			\textbf{Step 1}: Update output distritbution in \eqref{updatephats}  \;
			\textbf{Step 2}: Update $Q_\mathsf{\hat{S}|\tilde{S}}(\hat{s}|\tilde{s})$ in \eqref{updatepsss} \; }
		\textbf{Step 3}: Collect data $ \mathcal{I} \gets \big(D_c^{(\lambda_s)}, I^{(\lambda_s)}(\mathsf{\tilde{S};\hat{S}}) \big)$ \;  } 
	\textbf{Step 4}: Construct $D(R)$ function by data fit in $\mathcal{I}$  \;
	\KwOut{Communication distortion in \eqref{Dcopt}.}
\end{algorithm}

$\bullet$ \textbf{Communication distortion.}

In this sub-problem, we aim to evaluate the minimum communication distortion $D_c^{(\mu)}$ achieved under the source distribution $P_\mathsf{\tilde{S}}^{(\mu)}(\tilde{s})$ and the channel capacity $I^{(\mu)}(\Rv{X};\Rv{Y})$. This scenario corresponds to a typical rate-distortion problem in lossy data transmission. The distortion-rate curve can be obtained by the optimization problem \cite[Chapter 9]{yeung2008information}   
\begin{equation} \label{RDcom}
\begin{aligned}
	\mathop { \text{min} }\limits_{Q_\mathsf{\hat{S}|\tilde{S}}(\hat{s}|\tilde{s})} \kern 5pt I(\mathsf{\tilde{S};\hat{S}}) - \lambda_s D_c    
\end{aligned} 
\end{equation}
where $\lambda_s$ represents the slope of the distortion-rate curve. The conventional BA algorithm can be leveraged to solve problem 
\eqref{RDcom}. By following a similar procedure in sub-problem 1, the optimal solution can be calculated by
\begin{equation} \label{updatepsss}
Q_\mathsf{\hat{S}|\tilde{S}}(\hat{s}|\tilde{s}) = \frac{P_\mathsf{\hat{S}}(\hat{s})e^{\lambda_s d(\tilde{s},\hat{s})}}{\sum_{\hat{s}}P_\mathsf{\hat{S}}(\hat{s}')e^{\lambda_s d(\tilde{s},\hat{s}')}}.
\end{equation}   
The unknown nuisance term $\Prob{\hat{S}}{\hat{s}}$ can be updated by
\begin{equation} \label{updatephats}
P_\mathsf{\hat{S}}(\hat{s}) = \sum_{\tilde{s}} P_\mathsf{\tilde{S}}(\tilde{s})Q_\mathsf{\hat{S}|\tilde{S}}(\hat{s}|\tilde{s}).
\end{equation} 
By setting an initial distribution $\Claw{\hat{S}|\tilde{S}}{\hat{s}|\tilde{s}}$, the optimal solution of \eqref{RDcom} can be obtained by updating \eqref{updatephats} and \eqref{updatepsss} iteratively. The detailed derivations can be found in \cite{yeung2008information}. 

For a given slop $\lambda_s$, the tangent point $\big( D^{(\lambda_s)}_c, I^{(\lambda_s)}(\mathsf{\tilde{S};\hat{S}}) \big)$ of the distortion-rate curve can be obtained by solving problem \eqref{RDcom}. We may collect a series of distortion-rate data points to fit a distortion-rate function by varying the slop $\lambda_s$. Thus, the communication distortion constrained by a given channel capacity is attained by   
\begin{equation} \label{Dcopt}
D_c^{(u)} = D_\text{IT}(I^{(u)}(\Rv{X};\Rv{Y})),
\end{equation}    
where $D_\text{IT}(\cdot)$ represents the distortion-rate function. We summarize the communication distortion computation algorithm into Algorithm \ref{alg2}.

$\bullet$ \textbf{\ac{sdas} distortion.} 

In the outer loop, we may collect a set $\mathcal{D}$ of the \ac{sdas} distortion $D^{(\mu)}=D_s^{(u)}+D_c^{(u)}$ and the associated channel input distribution $P_\mathsf{X}^{(\mu)}(x)$ by repeating the above steps with various penalty factor $\mu$. Therefore, the optimal channel distribution that minimizes the \ac{sdas} distortion can be found by     
\begin{equation} \label{optsolution}
	D^{(\mu^\star)} = \mathop { \arg \min } \limits_{D \in \mathcal{D}}  \kern 2pt D, \kern 5pt  P_\mathsf{X}^{(\star)}(x) =  P_\mathsf{X}^{(\mu^\star)}(x).
\end{equation}     
The proposed two-step BA-based optimal search algorithm is summarized in Algorithm \ref{alg3}. 

\begin{algorithm}[!t]
	\caption{Two-step BA-based Optimal Search}
	\label{alg3}
	\KwIn{The set of penalty factor $\mathcal{U} \subset [0, \mu_{\text{max}}]$.}
	\For{$u \in \mathcal{U}$}{
		\textbf{Step 1}: Obtain $P^{(\mu)}_\mathsf{X}(x)$, $D^{(\mu)_s}$, and $I^{(u)}(\Rv{X};\Rv{Y})$ through Algorithm \ref{alg1} \;
		\textbf{Step 2}: Calculate estimate distribution $P^{(\mu)}_\mathsf{\tilde{S}}(\tilde{s})$ \;
		\textbf{Step 3}: Obtain $D_c^{(\mu)}$ through Algorithm \ref{alg2} \;    
		\textbf{Step 4}: $\mathcal{D} \gets D^{(\mu)}=D_s^{(\mu)} +D_c^{(\mu)}$ \;
	}
	\KwOut{Optimal solution in \eqref{optsolution}.}
\end{algorithm}

\section{Waveform Design for Gaussian Channel Input} \label{WaveformDesign}

The previous section illustrates that using Gaussian channel input may not be optimal for the \ac{sdas} systems. Despite this limitation, the assumption of Gaussian signals is prevalent in communications research literature. In this section, we introduce an ISAC waveform design scheme specifically developed for multi-input multi-output (MIMO) systems, while adhering to the framework of Gaussian signaling. To proceed, we need to derive the explicit expressions of the variables in $\mathcal{P}_0$ for the MIMO case. 
      
\subsection{System Model}
The widely employed signal models of the \ac{snc} processes for the MIMO systems are expressed by     
\begin{equation}\label{glm1}
\mathbf{Z}=\mathbf{H}_s\mathbf{X}+\mathbf{N}_s, \kern 5pt  \mathbf{Y}=\mathbf{H}_c\mathbf{X}+\mathbf{N}_c,
\end{equation}  
where $\mathbf{X}\in \mathbb{C}^{N_t \times T}$, $\mathbf{Z} \in \mathbb{C}^{M_s \times T}$, and $\mathbf{Y} \in \mathbb{C}^{M_c \times T}$ represent the transmitting ISAC signal, the received signals at the BS (encoder) and the end-user (decoder), respectively; $\mathbf{H}_s$ and $\mathbf{H}_c$ denote the \ac{snc} channel state information (CSI) matrices; $\mathbf{N}_s$ and $\mathbf{N}_c$ are the channel noises whose entries follow the complex Gaussian distribution with $\mathcal{CN}(0,\sigma_s^2)$ and $\mathcal{CN}(0,\sigma_c^2)$; Finally, $N_t$, $M_s$, $M_c$, and $T$ denote the numbers of Tx antennas, \ac{snc} receiver (Rx) antennas, and the transmitting symbols, respectively.

We focus on the sensing task of \ac{trm} estimation, where the to-be-estimate parameters can be written by $\mathbf{s} = \text{vec}(\mathbf{H}_s)$. After vectorization operations, the sensing signal observed at the BS can be recast by         
\begin{equation}\label{vecs}
\mathbf{z}=\left(\mathbf{I}_{M_s} \otimes \mathbf{X}^H \right)\mathbf{s}+\mathbf{n}_s,
\end{equation}  
with $\mathbf{z}=\text{vec}(\mathbf{Z}^H)$, and $\mathbf{n}_s = \text{vec}(\mathbf{N}_s^H)$. The BS yields an estimate of the \ac{trm}, namely, $\tilde{\mathbf{s}}$, from the observations $\mathbf{z}$, then transmits it to the user through communication channel $\mathbf{H}_c$. Subsequently, the user can reconstruct $\hat{\mathbf{s}}$ from the communication received data $\mathbf{Y}$. Here, we adopt the quadratic distortion metric (i.e., \ac{mse}), which meets the separable condition in \eqref{qudraticmtric}. Before proceeding the ISAC waveform design, we make the following assumptions.       

\textit{\textbf{Assumption 1}}: The ISAC waveform $\mathbf{X}$ follows complex Gaussian distribution with $\mathcal{CN}(0, \mathbf{R}_x)$, where $\mathbf{R}_x \in \mathbb{C}^{N_t \times N_t}$ is the covariance matrix.          

\textit{\textbf{Assumption 2}}: The \ac{trm} vector $\mathbf{s}$ follows complex Gaussian distribution $\mathcal{CN}(0,\mathbf{I}_{M_s} \otimes \bm{\Sigma}_s)$, where $\bm{\Sigma}_s \in \mathbb{C}^{N_t \times N_t} $ denotes the covariance matrix of each column of $\mathbf{H}_s$.\footnote{This assumption corresponds to the scenario that the Rx antennas for sensing are sufficiently separated so that the correlations among the rows of $\mathbf{H}_s$ can be ignored \cite{8579200}. Here, we specify this Kronecker structure in TRM covariance matrix to simplify the expression of sensing distortion \cite{5272481}. To avoid the deviation of our core contribution in this paper, we will leave the general TRM covariance matrix cases for future research.} 

The above assumptions enable us to derive the explicit expressions of the \ac{snc} distortions, the estimate distribution $\tilde{\mathbf{s}}$, and the communication channel capacity \ac{wrt} the ISAC waveform matrix $\mathbf{X}$.  
   
1) \textit{Sensing distortion $D_s$:} For the Gaussian linear model \eqref{vecs}, the optimal estimator \eqref{escost}, i.e., the well-known MMSE estimator, is leveraged to estimate $\tilde{\mathbf{s}}$ by 
\begin{equation}\label{estimates}
\tilde{\mathbf{s}} = \Big(\mathbf{I}_{M_s} \otimes \Big(\bm{\Sigma}_s\mathbf{X}\left(\mathbf{X}^H \bm{\Sigma}_s \mathbf{X} + \sigma^2_s\mathbf{I}_T\right)^{-1}\Big)\Big)\mathbf{z},
\end{equation}
Furthermore, the sensing distortion is attained by \cite{BookEstimationTheory}  
\begin{equation}\label{mmse1}
	D_s (\mathbf{R}_x) = \mathbb{E}\Big[\|\mathbf{s}-\tilde{\mathbf{s}}\|^2\Big] = M_s \tr{\Big( \frac{1}{\sigma^2_s}\mathbf{R}_x + \bm{\Sigma}_s^{-1} \Big)^{-1}}  .
\end{equation}

2) \textit{Rate distortion function $R(D_c)$:} The estimate $\tilde{\mathbf{s}}$ in \eqref{estimates} follows the complex Gaussian distribution with $\mathcal{CN}(\mathbf{0},\mathbf{I}_{M_s} \otimes \mathbf{R}_{\tilde{s}})$, with the covariance matrix of \cite{BookEstimationTheory} 
\begin{equation}\label{reta}
	\mathbf{R}_{\tilde{\mathbf{s}}} = \bm{\Sigma}_s - \Big( \frac{1}{\sigma^2_s}\mathbf{R}_x + \bm{\Sigma}_s^{-1} \Big)^{-1}.
\end{equation}  
Let us temporarily omit the Kronecker product and focus on a source with Gaussian distribution $\mathcal{CN}(\mathbf{0}, \mathbf{R}_{\tilde{s}})$. In light of the results in \cite{9844779}, the rate-distortion function is equivalent to the objective of the following optimization problem     
\begin{equation} \label{rdgua}
\begin{aligned}
 R(D_c) = \opmin{\mathbf{D} \in \mathbb{S}_{N_t}^+} & \log \left( \frac{\det(\mathbf{R}_{\tilde{s}})}{\det(\mathbf{D})} \right) \\
 \text{subject to} & \kern 5pt \mathbf{D} \preceq \mathbf{R}_{\tilde{s}}, \kern 2pt \tr{\mathbf{D}} \le D_c,
\end{aligned}
\end{equation}
where $\mathbb{S}_{N_t}^+$ denotes the set of all $N_t \times N_t$ positive definite matrices. Denote $\mathbf{D}^\star$ as the solution of \eqref{rdgua}, the communication distortion $D_c$ and the required rate $R(D_c)$ can be obtained by 
\begin{equation}
D_c = M_s\tr{\mathbf{D}^\star}, \kern 5pt R(D_c) = M_s \log \left( \frac{\det(\mathbf{R}_{\tilde{s}})}{\det(\mathbf{D}^\star)} \right),    
\end{equation}  
where $M_s$ is due to the Kronecker product. The derivations of \eqref{rdgua} can be found in the proof of \cite[Theorem 2]{9844779} \footnote{Our result is a straightforward corollary with the semantic distortion equal to zero in Theorem 2.}.

3) \textit{Channel Capacity:} The MIMO Gaussian channel capacity can be expressed by
\begin{equation} \label{cap}
C(\mathbf{R}_x) = \log \det \left(\frac{1}{\sigma^2_c}\mathbf{H}_c\mathbf{R}_x\mathbf{H}_c^H+\mathbf{I}_{M_c} \right).
\end{equation}      

\begin{algorithm}[!t]
	\caption{SCA-based Waveform Design Algorithm}
	\label{alg4}
	\KwIn{An initial covariance matrix $\mathbf{R}_0$, $K$.}
	\For{$k \le K$}{
		\textbf{Step 1}: Obtain optimal $\mathbf{R}^{\star}_x$ by solving $\mathcal{P}_3$ \;
		\textbf{Step 2}: $\mathbf{R}_0 = \mathbf{R}^{\star}_x $ \;
	}
	\KwOut{Optimal solution $\mathbf{R}^\star_x$.}
\end{algorithm}

\subsection{ISAC Waveform Design}
Observe that the expressions \eqref{mmse1} to \eqref{cap} are all related to the covariance matrix $\mathbf{R}_x$ rather than the specific waveform matrix $\mathbf{X}$. Consequently, our ISAC waveform design problem can be transformed into determining the optimal covariance matrix that minimizes the \ac{sdas} distortion. Regarding the system resource constraint, we only consider the transmit power at the BS. By substituting the expressions into the original problem $\mathcal{P}_1$ and introducing an auxiliary variable of positive definite matrix $\mathbf{D}$, the ISAC waveform design problem can be reformulated by  
\begin{equation} \label{p2}
\mathcal{P}_2 \left\{ \begin{aligned}
\opmin{\mathbf{R}_x,\mathbf{D} \in \mathbb{S}_{N_t}^+} & D_s(\mathbf{R}_x) + \tr{\mathbf{D}}  \\
\text{subject to} & \kern 5pt  C(\mathbf{R}_x) - M_s \log \left( \frac{\det(\mathbf{R}_{\tilde{s}})}{\det(\mathbf{D})} \right) \ge 0,   \\
& \kern 5pt   \mathbf{D} \preceq \mathbf{R}_{\tilde{s}}, \kern 2pt \mathbf{R}_x \succeq 0, \kern 2pt \tr{\mathbf{R}_x} \le P_T,
\end{aligned} \right.
\end{equation}   
where $P_T$ is the power budget. $\mathcal{P}_2$ is non-convex due to the nonlinear intermediate variables $\log\det (\mathbf{R}_{\tilde{s}})$ and $\mathbf{R}_{\tilde{s}}$. To relax $\mathcal{P}_2$ into a convex problem, we employ an SCA technique based on Taylor series expansion. To be specified, for any given point $\mathbf{R}_0$, the matrix $\mathbf{R}_{\tilde{s}}$ can be approximated by       
\begin{equation} \label{Taylor1}
\begin{aligned}
\mathbf{R}_{\tilde{s}} \simeq \bm{\Sigma}_s - \mathbf{P} + \frac{1}{\sigma_s} \mathbf{P} (\mathbf{R}_x - \mathbf{R}_0) \mathbf{P} \triangleq \widetilde{\mathbf{R}}_{\tilde{s}},   
\end{aligned}
\end{equation}
with the constant matrix defined by
\begin{equation}
\mathbf{P}= \Big( \frac{1}{\sigma^2_s}\mathbf{R}_0 + \bm{\Sigma}_s^{-1} \Big)^{-1}.  \notag
\end{equation} 
Furthermore, we have 
\begin{equation} \label{Taylor2}
\begin{aligned}
& \log \det (\mathbf{R}_{\tilde{s}}) \simeq \log\det \left( \bm{\Sigma}_s - \mathbf{P} \right) \\
  & \kern 10 pt + \tr{\frac{1}{\sigma_s^2} (\bm{\Sigma}_s - \mathbf{P})^{-1} \mathbf{P} (\mathbf{R}_x - \mathbf{R}_0) \mathbf{P}} \triangleq f(\mathbf{R}_x).
\end{aligned}
\end{equation}
Note that $\widetilde{\mathbf{R}}_{\tilde{s}}$ and $f(\mathbf{R}_x)$ are both the linear functions \ac{wrt} variable $\mathbf{R}_x$ with a given point $\mathbf{R}_0$. By substituting \eqref{Taylor1} and \eqref{Taylor2} into $\mathcal{P}_2$, the problem can be relaxed into
\begin{equation} \label{p3}
	\mathcal{P}_3 \left\{ \begin{aligned}
		\opmin{\mathbf{R}_x,\mathbf{D} \in \mathbb{S}_{N_t}^+} & D_s(\mathbf{R}_x) + \tr{\mathbf{D}}  \\
		\text{subject to} & \kern 5pt  C(\mathbf{R}_x) - M_s \left( f(\mathbf{R}_x) - \log \det (\mathbf{D}) \right)  \ge 0,   \\
		& \kern 5pt   \widetilde{\mathbf{R}}_{\tilde{s}} - \mathbf{D} \succeq 0, \kern 2pt \mathbf{R}_x \succeq 0, \kern 2pt \tr{\mathbf{R}_x} \le P_T.
	\end{aligned} \right.
\end{equation}
For a given point $\mathbf{R}_0$, $\mathcal{P}_3$ is convex since both the objective and constraints are either convex or linear. Therefore, it can be efficiently solved by using the off-the-shelf CVX toolbox \cite{grantcvx}. For an initial covariance matrix, e.g., $\mathbf{R}_0 = P_T/N_t\mathbf{I}$, the non-convex problem $\mathcal{P}_2$ can be addressed by iteratively computing the optimal solution $\mathbf{R}^{\star}_x$ of $\mathcal{P}_3$ and updating $\mathbf{R}_0 = \mathbf{R}^{\star}_x$ until convergence. The procedure is outlined in Algorithm \ref{alg4}.

\section{Simulation Results} \label{simulationR}

The simulation results are divided into two sections. The first section evaluates the optimal distribution for the ISAC channel input in scalar cases, demonstrating the effectiveness of the proposed MSST and the two-step BA-based optimal search algorithm. Moreover, the unique distortion-capacity (D-C) curves of the \ac{sdas} system are provided to illustrate the performance tradeoff between \ac{snc} processes. The second section presents the results of the optimal waveform design for the Gaussian MIMO channel, showing its superiority over the \ac{snc}-optimal schemes.

\subsection{Optimal Distribution for ISAC Channel Input}

In this subsection, we aim to reveal the unique performance tradeoff between \ac{snc} processes. The simulation parameters are set as follows. The prior distribution of the target's sate is assumed to be $\Prob{S}{s} \sim \mathcal{CN}(0,1)$, i.e, $\nu_s^2=1$. We define the normalized signal-to-noise ratios (SNRs) for the \ac{snc} channels as $10\log(1/\sigma^2)$ dB and set the power budget to be $B=5$. Unless otherwise specified, the \ac{snc} channel SNRs are set to be $\text{SNR}_s=\text{SNR}_c=0$ dB in this subsection. To improve the computational efficiency, the weighted factor $\mu$ takes values from the sets $[0,1]$, $[1,5]$, and $[5,30]$ with non-uniform spacings of $0.1$, $0.5$, and $5$, respectively.    
  
\begin{figure}[!t]
	\centering
	\includegraphics[width=3in]{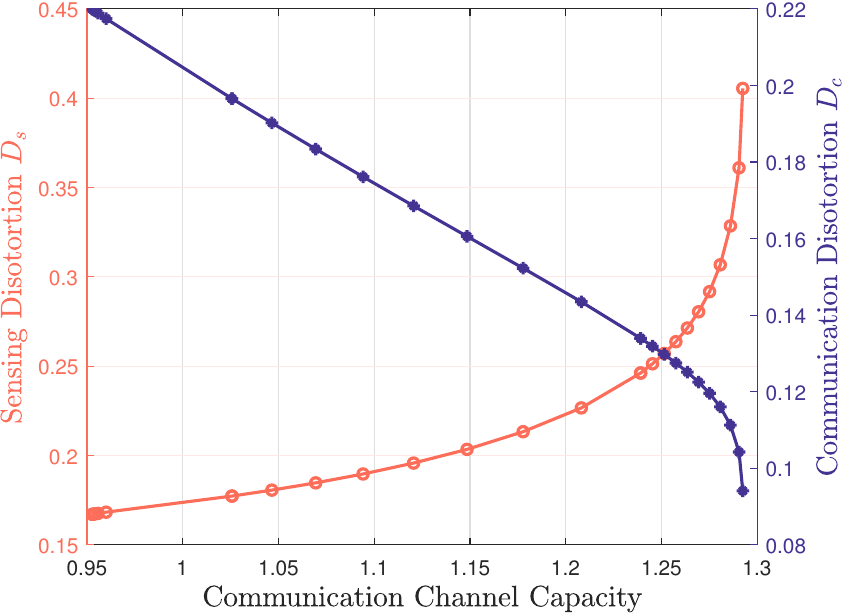}
	\caption{The changes of the \ac{snc} distortions with various factor $\mu$.}
	\label{DSDC}
\end{figure}

\begin{figure}[!t]
	\centering
	\includegraphics[width=3in]{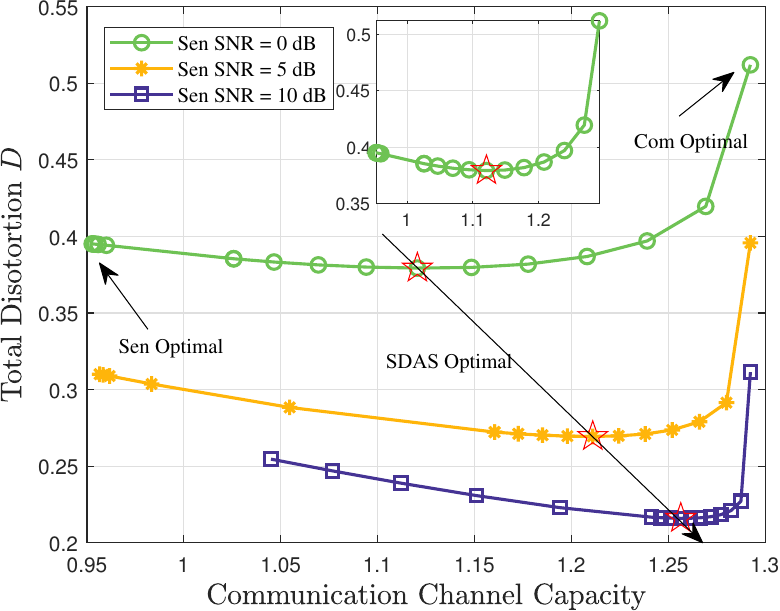}
	\caption{The SAS distortion versus the communication channel capacity.}
	\label{CD_Curve}
\end{figure}

\begin{figure}[!t]
	\centering
	\includegraphics[width=2.8in]{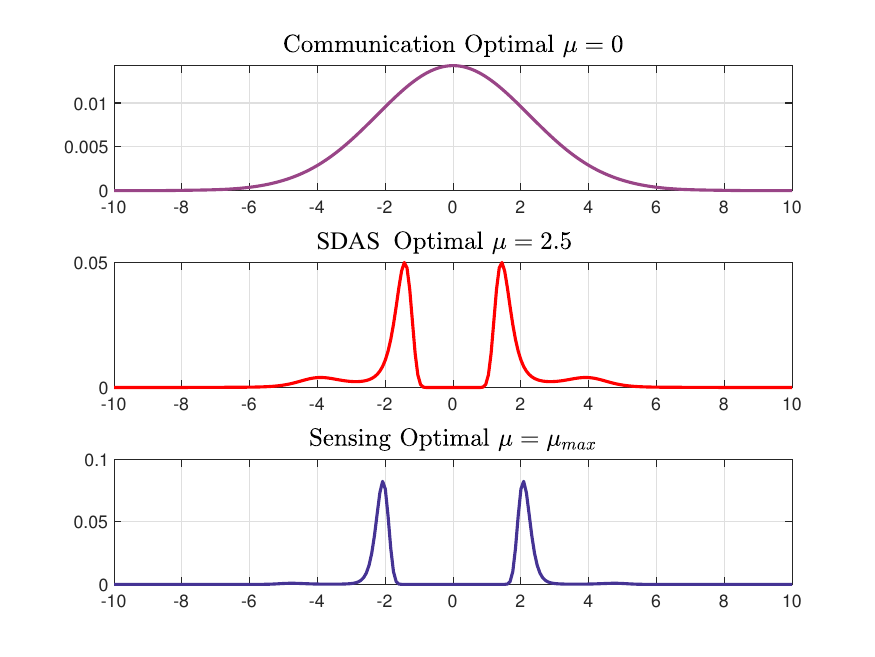}
	\caption{The channel input distribution of \ac{snc}-optimal and SAS-optimal.}
	\label{OptimalPDF}
\end{figure}

\subsubsection{The Distortion-Capacity (D-C) curves and the optimal distribution of ISAC channel input}
Fig. \ref{DSDC} presents the D-C curves for both \ac{snc} processes. As expected, sensing distortion $D_s$ increases as the values of $\mu$ decrease, implying that the estimation performance at the BS becomes less significant. Conversely, the communication channel capacity increases accordingly, leading to a reduction in communication distortion $D_c$. Consequently, these opposing trends in \ac{snc} distortions result in an uncertain \ac{sdas} distortion. In Fig. \ref{CD_Curve}, we show the resultant D-C curve for the \ac{sdas} systems with different sensing SNRs. A \ac{sdas}-optimal point that minimizes the \ac{sdas} distortion is evident, highlighting a unique phenomenon in \ac{sdas} systems compared to the existing studies \cite{9785593,10147248}.

Fig. \ref{OptimalPDF} depicts the specific channel input distributions corresponding to three special points in Fig. \ref{CD_Curve}, namely, the \ac{snc}-optimal and \ac{sdas}-optimal points. The channel capacity is maximized without the sensing constraint when $\mu = 0$. It is well-known that Gaussian channel input achieves the maximum MI for the Gaussian channels. By contrast, at the sensing-optimal point, as $\mu$ becomes sufficiently large, the channel input exhibits a 2-ary pulse amplitude modulation. These results are consistent with the findings in \cite{8437621}. However, the unique \ac{sdas}-optimal distribution, presenting a compromise between the \ac{snc}-optimal distributions, is attained at $\mu = 2.5$.                        

\begin{figure}[!t]
	\centering
	\includegraphics[width=2.93in]{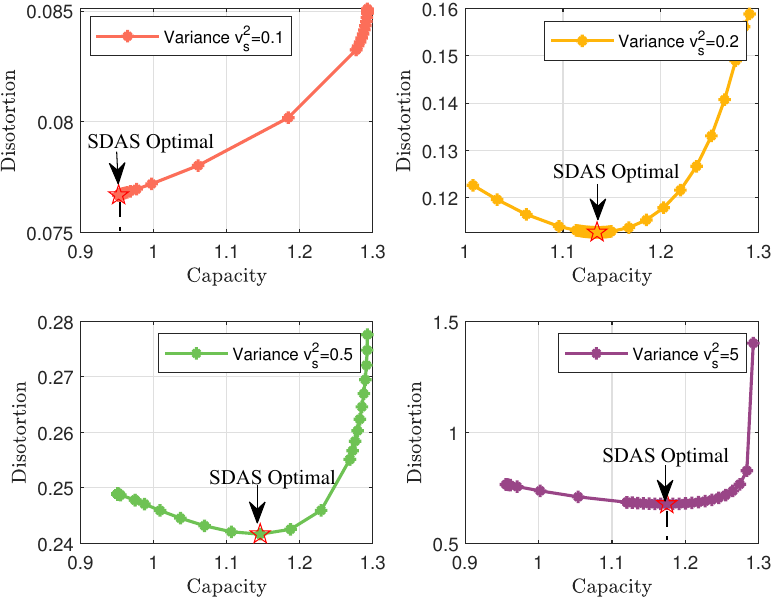}
	\caption{The distortion-capacity curves with various state variances.}
	\label{CDVSV}
\end{figure}

\begin{figure}[!t]
	\centering
	\includegraphics[width=2.93in]{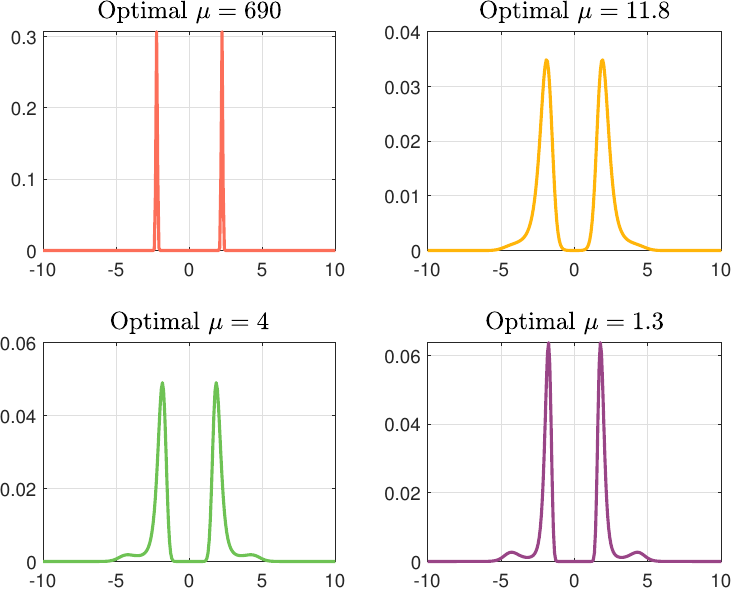}
	\caption{The \ac{sdas}-optimal distribution for various state variances.}
	\label{distri_CD}
\end{figure}

\subsubsection{The impact of the target's prior variance}
Subsequently, we investigate the impact of the target's state variance by selecting the values of $\nu_s^2 \in \{0.1,0.2,0.5,5\}$. Due to the significant differences in the order of magnitude of the values for MI and sensing distortions in \eqref{maxCMI2}, we adjust the interval of $\mu$ for each state variance rather than using a fixed interval. The essential purpose of the \ac{sdas} systems is to reduce the uncertainty of the target's prior information, specifically the variance of the state's prior distribution. 

In Fig. \ref{CDVSV}, each point represents a lower \ac{sdas} distortion compared to the prior variance, even at the \ac{snc}-optimal points. This reduction is attributed to the resource multiplexing gain achieved through the dual-functional signaling strategy. Additionally, we observe that the \ac{sdas}-optimal point gradually shifts from sensing-optimal to communication-optimal as variance increases. A small variance indicates relatively accurate prior information about the target's state, necessitating an ISAC channel input with strong sensing capabilities for better accuracy performance. Conversely, as variance increases, greater emphasis on communication performance is required to convey more information.     

We provide the detailed distributions of the \ac{sdas}-optimal channel input for different state variances. For a small variance (e.g., $\nu_s^2=0.1$), the sensing distortion becomes small accordingly. It is at least less than the prior variance; otherwise, the user can `guess' the state information in terms of the prior distribution. Therefore, a large weighted factor is required, e.g., $\mu = 690$, to balance the values of MI and sensing distortion in \eqref{maxCMI2}. In this case, the \ac{sdas}-optimal distribution aligns with the sensing-optimal distribution, specifically a standard 2-ary pulse amplitude modulation. As anticipated, the optimal $\mu$ values decrease with increasing state variance, and the \ac{sdas}-optimal distribution changes accordingly.

\subsection{Optimal Waveform Design for Gaussian Signal}

In this subsection, we demonstrate the effectiveness of the proposed ISAC waveform design scheme for Gaussian signaling. The general system setups are as follows. The communication channel is modeled by Rayleigh fading, where each entry of $\mathbf{H}_c$ obeys the standard complex Gaussian distribution. All results are obtained by the average of $100$ Monte Carlo trials and the average distortion in each Monte Carlo trial is calculated by $D/(M_sN_t)$. 
          
\begin{figure}[!t]
	\centering
	\includegraphics[width=3in]{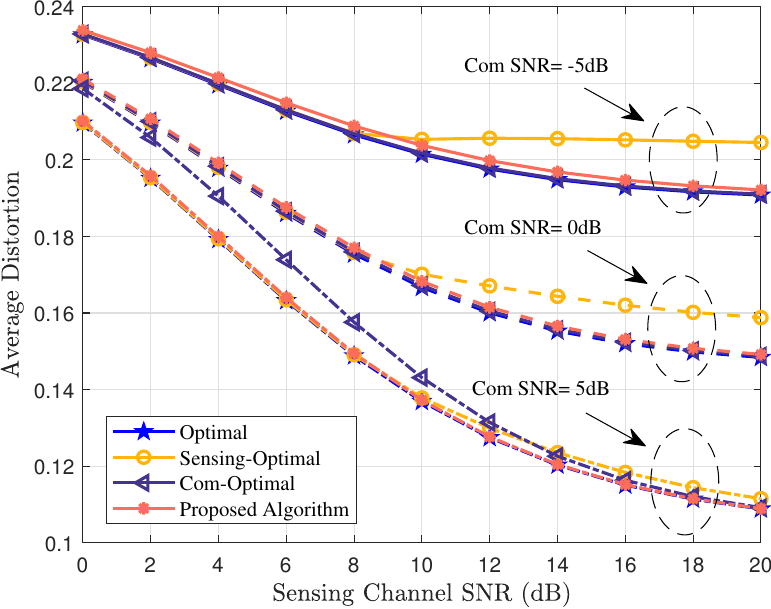}
	\caption{The average distortion versus sensing channel SNRs for various methods, $a_1 = 0.4, a_2=0.1$, $N_t = M_c = M_s =2$.}
	\label{D_SNRs_N2}
\end{figure}

\subsubsection{Waveform design for a 2-D special case}
We begin with a two-dimensional (2-D) special case where the state covariance matrix, $\bm{\Sigma_s} = \text{diag}\{a_1,a_2\}$, is assumed to be diagonal. In this scenario, the global optimal solution of problem $\mathcal{P}_2$ can be attained through a 2-D exhaustive search, as outlined in \cite{10845869}. Thus, this brute force solution \footnote{The optimal eigenspace of $\mathbf{R}_x$ must align with the communication eigenspace of $\mathbf{H}_c^H\mathbf{H}_c$ as the sensing eigenspace is an arbitrary unitary matrix. Thus, the original problem can be reduced to finding two optimal eigenvalues over 2-D grids, with the sum of each 2-D point equal to $P_T$.} can serve as a benchmark for comparing the superiority of our method. Accordingly, the numbers of the Tx and \ac{snc} Rx antennas are set by $N_t=M_c=M_s=2$.

Fig. \ref{D_SNRs_N2} illustrates the impact of sensing channel quality on the achievable average distortion. The sensing SNRs vary from $0$ dB to $20$ dB with a spcing $2$ dB. Besides the global optimal benchmark, the \ac{snc} optimal waveform schemes refer to the solutions that minimize the MMSE in \eqref{mmse1} and maximize the channel capacity in \eqref{cap}, respectively. At a glance, we observe that the achievable average distortion decreases with increasing sensing channel SNRs. Furthermore, the sensing-optimal scheme is preferable for acquiring accurate target information when the communication channel is sufficiently good at $\text{SNR}_c=5$ dB. However, the sensing-optimal scheme may cause significant performance degradation as the communication SNR decreases. Particularly, the sensing-optimal curve even tends to be flat at $\text{SNR}_c=-5$ dB, implying that a poor communication channel severely limits performance improvement. This clearly highlights the unique tadeoff between \ac{snc} processes in \ac{sdas} systems. Finally, it is worth noting that the proposed algorithm achieves satisfactory performance compared to the global optimal solution.                

 \begin{figure}[!t]
 	\centering
 	\includegraphics[width=3in]{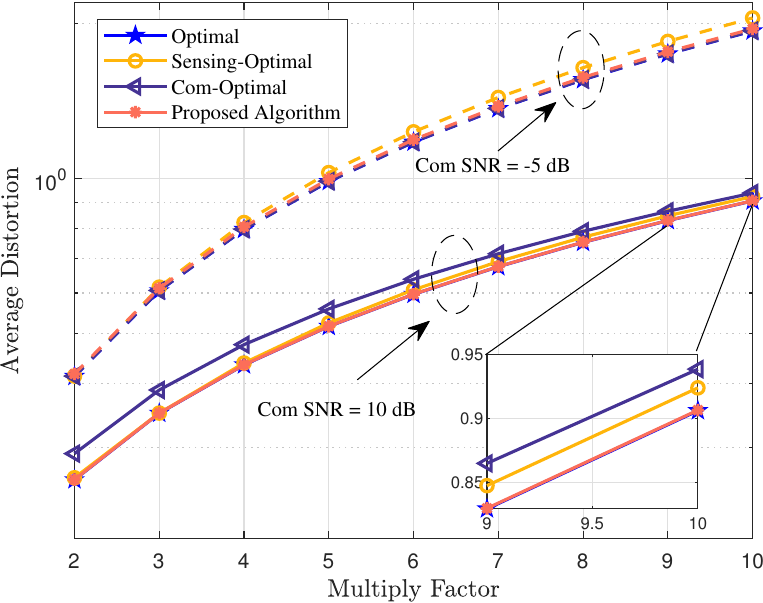}
 	\caption{The impact of the state variances on the SAS performance, $ \text{SNR}_s = 5$ dB, $N_t = M_c = M_s =2$.}
 	\label{MultiFactor}
 \end{figure}

In Fig. \ref{MultiFactor}, we investigate the impact of state variances on the achievable average distortion in the MIMO systems. The state variances are scaled by a factor ranging from $2$ to $10$. Two interesting observations emerge. 1) For small state variances, the global optimal scheme coincides with the sensing-optimal scheme at $\text{SNR}_c = 10$ dB. However, both the \ac{snc}-optimal schemes experience performance degradation as state variances increase. 2) When the communication channel quality is insufficient at $\text{SNR}_c = -5$ dB, the global optimal scheme tends towards the communication-optimal scheme to enhance communication capabilities. As state variances increase, the performance degradation of the sensing-optimal scheme becomes more significant. Nevertheless, our proposed algorithm can effectively balance the \ac{snc} processes, resulting in improved average distortion.                   

\begin{figure}[!t]
	\centering
	\includegraphics[width=3in]{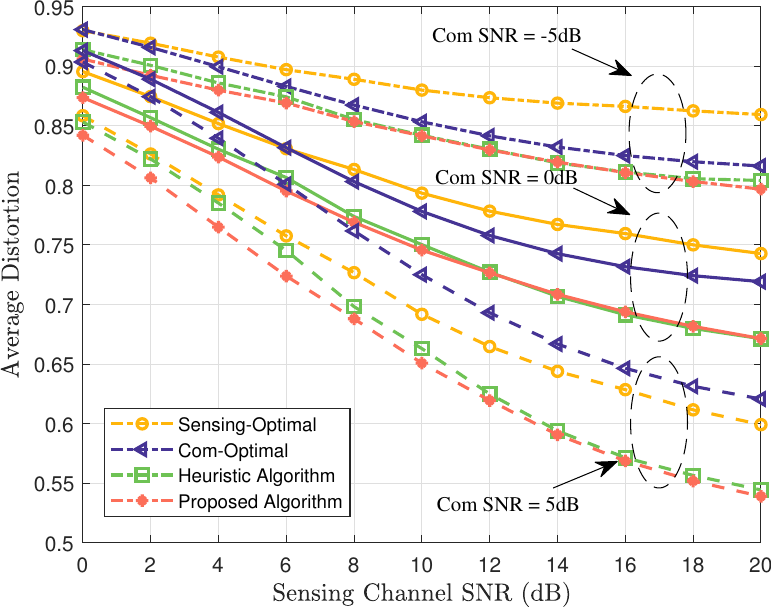}
	\caption{The average distortion versus sensing channel SNRs for various methods, $N_t=10, M_s=2, M_c=5$.}
	\label{MultiAntenna_SenSNR}
\end{figure}

\subsubsection{Waveform design for general MIMO cases}
Subsequently, we extend the results to high-dimensional matrix scenarios, where the number of Tx antennas is $N_t = 10$, and the numbers of \ac{snc} Rx antennas are $M_s = 2$ and $M_c = 5$, respectively. The state covariance matrix $\bm{\Sigma}_s$ is randomly generated as an $N_t \times N_t$ Hermitian matrix. It is worth noting that obtaining the global optimal solution is challenging due to the highly coupled \ac{snc} processes. For comparison purposes, we provide the results of the Heuristic algorithm proposed in \cite{10845869}, whose formulation is given by
\begin{equation}
\begin{aligned}
\opmax{\mathbf{R}_x} & \kern 5pt \beta I(\mathbf{Y};\mathbf{X}|\mathbf{H}_c) + (1-\beta) I(\mathbf{Z};\mathbf{H}_s|\mathbf{X})  \\
\text{subject to} & \kern 5pt  \tr{\mathbf{R}_x} \le P_T,
\end{aligned} 
\end{equation} 
where $I(\mathbf{Z};\mathbf{H}_s|\mathbf{X}) = \log \det \left(\frac{1}{\sigma^2_s}\bm{\Sigma}_s\mathbf{R}_x+\mathbf{I}_{N_t} \right)$ is defined by sensing MI and $I(\mathbf{Y};\mathbf{X}|\mathbf{H}_c)$ is defined in \eqref{cap}. The weighted factor $\beta$ takes values over the interval $[0,1]$ with $L$ grid points. In our simulations, $L$ is set to be $11$.   
       
Fig. \ref{MultiAntenna_SenSNR} illustrates the average distortions versus sensing channel SNRs for various waveform design schemes. It can be observed that the proposed algorithm outperforms its counterparts. Similar trends to those shown in Fig. \ref{D_SNRs_N2} for the 2-D case are also noticeable. Besides, we can observe that the proposed ISAC waveform design achieves significantly greater performance gains in high-dimensional scenarios.  

\begin{figure}[!t]
	\centering
	\includegraphics[width=3in]{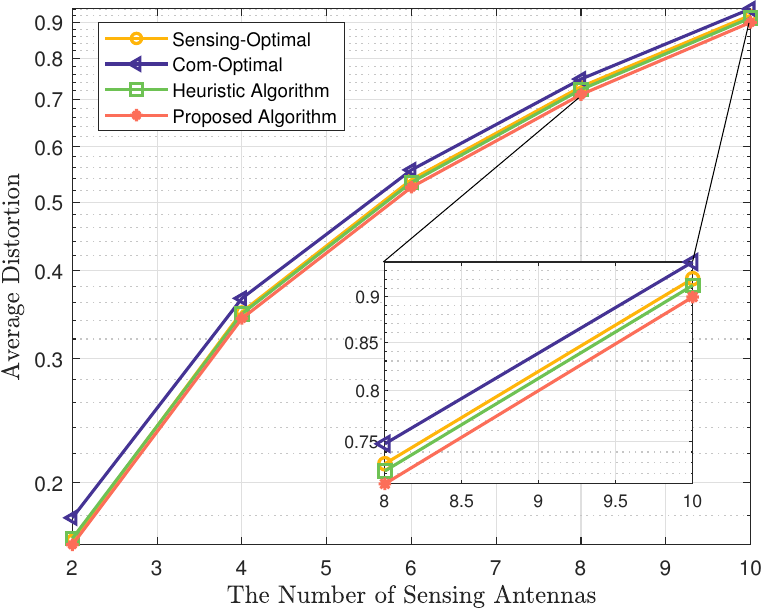}
	\caption{The average distortion versus the number of sensing Rx antennas, $N_t=10, M_c=5$, $\text{SNR}_s = 0$ dB, $\text{SNR}_c = 10$ dB.}
	\label{Distortion_SenAntennas}
\end{figure}

\begin{figure}[!t]
	\centering
	\includegraphics[width=3in]{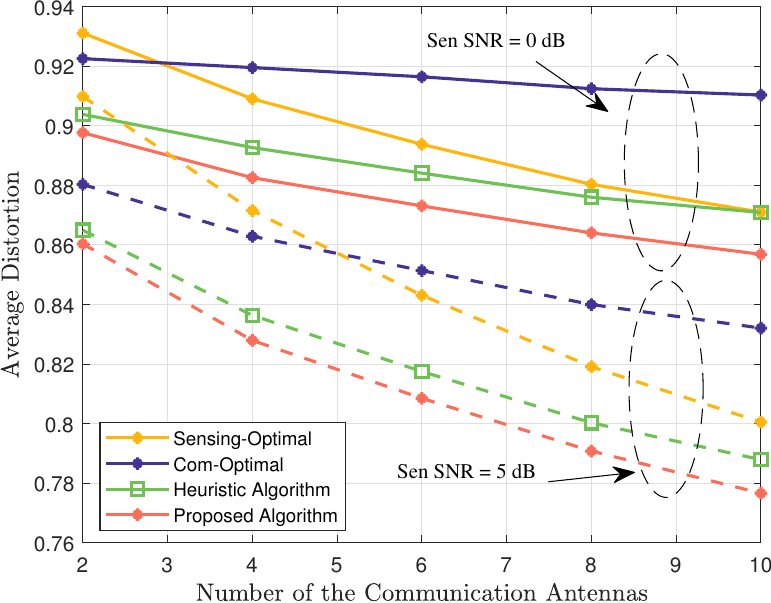}
	\caption{The average distortion versus the number of communication Rx antennas, $N_t=10, M_s=2$, $\text{SNR}_c = 0$ dB.}
	\label{Distortion_ComAntenna}
\end{figure}

Fig. \ref{Distortion_SenAntennas} and Fig. \ref{Distortion_ComAntenna} demonstrate the influence of \ac{snc} Rx antennas on average distortion across various waveform design schemes. Fig. \ref{Distortion_SenAntennas} exhibits a linear increase in average distortion with the addition of sensing Rx antennas. The linearity feature consistent with the Kronecker product form in assumption 2. Moreover, regarding the TRM estimation, an increase in sensing Rx antennas necessitates more channel information to be estimated, leading to a larger average distortion.

In contrast, Fig. \ref{Distortion_ComAntenna} shows a declining trend in average distortion as the number of communication Rx antennas increases. This performance improvement is attributed to enhanced communication channel capacity facilitated by multiple Rx antenna gain, thereby improving overall system performance. Additionally, under the communication-optimal scheme, there is a gradual performance degradation as channel capacity increases. This trend arises because improving sensing performance becomes critical in scenarios with excessive communication channel capacity. Once again, the proposed ISAC waveform design method is superior to the other counterparts in all scenarios.

\section{Conclusion} \label{Conclude}

In this paper, we delve into the dual-functional signaling strategy for simultaneous sensing data acquisition and sharing (SDAS) systems across three primary aspects. Firstly, we develop a modified source-channel separation theorem tailored (MSST) for \ac{sdas} systems under the condition of separable distortion metric. We elucidate the operational meaning of the proposed MSST from the perspective of coding theory. Secondly, we develop an input distribution optimization scheme for \ac{sdas} systems under scalar channels, which minimizes the \ac{sdas} distortion while adhering to the MSST constraint. We determine the optimal distribution, balancing between the conventional communication-optimal (Gaussian distribution) and sensing-optimal (2-ary pulse amplitude modulation) schemes. Thirdly, we propose an ISAC waveform design method for Gaussian signaling in MIMO \ac{sdas} systems, where a successive convex approximation algorithm is conceived to solve for the optimal covariance matrix. Simulation results show the unique performance tradeoff between the sensing and communication processes.  

%%%%%%
%% Appendix:
%% If needed a single appendix is created by
%%
%\appendix
%%
%% If several appendices are needed, then the command
%%
% \appendices
%%
%% in combination with further \section commands can be used.
%%%%%%
%\IEEEtriggeratref{4}

\bibliographystyle{IEEEtran}
\bibliography{IEEEabrv,JSTSP_Revised}

\end{document}